\documentclass[structabstract]{aa}  
\usepackage{graphicx}
\usepackage{txfonts}
\usepackage{sidecap}
\usepackage{comment}
\usepackage{color}
\usepackage{natbib} 
\usepackage{bm,xstring}
\bibpunct{(}{)}{;}{a}{}{,} 

\begin{document}
\title{Search and characterization of T-type planetary mass candidates in the $\sigma$ Orionis cluster}
\subtitle{}
\author{K. Pe\~na Ram\'irez\inst{1,2} \and M.R. Zapatero Osorio\inst{3} \and V.J.S. B\'ejar\inst{1,2} \and R.Rebolo\inst{1,2,4} \and G. Bihain\inst{5} }
\institute{Instituto de Astrof\'isica de Canarias (IAC), E-38205 La Laguna, Tenerife, Spain. \email{karla@iac.es} \and Departamento de Astrof\'isica, Universidad de La Laguna, E-38205 La Laguna, Tenerife, Spain. \and Centro de Astrobiolog\'ia (CSIC-INTA), E-28850 Torrej\'on de Ardoz, Madrid, Spain. \and Consejo Superior de Investigaciones Cient\'ificas (CSIC), Madrid, Spain. \and Leibniz Institute for Astrophysics Potsdam, An der Sternwarte 16, D-14482 Potsdam, Germany.\\}
\date{Received 2011 / Accepted 2011}

\abstract
{The proper characterization of the least massive population of the young $\sigma$\,Orionis star cluster is required to understand the form of the cluster mass function and its impact on our comprehension of the substellar formation processes. S\,Ori\,70 (T5.5\,$\pm$\,1) and 73, two T-type cluster member candidates, would have likely masses between 3 and 7 M$_{\rm Jup}$ if their age is 3 Myr. 
S\,Ori\,73 awaits confirmation of its methane atmosphere.}
{We aim to: i) confirm the presence of methane absorption in S\,Ori\,73 through methane imaging; ii) study S\,Ori\,70 and 73 cluster membership via photometric colors and accurate proper motion analysis; iii) perform a new search to identify additional T-type $\sigma$\,Orionis member candidates.}
{We obtained HAWK-I (VLT) $J$, $H$, and $CH_{\rm 4off}$ photometry of an area of 119.15 arcmin$^2$ in $\sigma$ Orionis down to $J_{\rm comp}$\,=\,21.7 and $H_{\rm comp}$\,=\,21 mag. S\,Ori\,70 and 73 are contained in the explored area. Near-infrared data were complemented with optical photometry using images acquired with OSIRIS (GTC) and VISTA as part of the VISTA Orion survey. Color-magnitude and color-color diagrams were constructed to characterize S\,Ori\,70 and 73 photometrically, and to identify new objects with methane absorption and masses below 7 M$_{\rm Jup}$. We derived proper motions by comparison of the new HAWK-I and VISTA images with published near-infrared data taken 3.4\,--\,7.9 yr ago.}
{S\,Ori\,73 has a red $H-CH_{\rm 4off}$ color indicating methane absorption in the $H$-band and a spectral type of T4\,$\pm$\,1. S\,Ori\,70 displays a redder methane color than S\,Ori\,73 in agreement with its latter spectral classification. Our proper motion measurements ($\mu_\alpha\,{\rm cos}\,\delta$\,=\,26.7\,$\pm$\,6.1, $\mu_\delta$\,=\,21.3\,$\pm$\,6.1 mas\,yr$^{-1}$ for S\,Ori\,70, and $\mu_\alpha\,{\rm cos}\,\delta$\,=\,46.7\,$\pm$\,4.9, $\mu_\delta$\,=\,$-$6.3\,$\pm$\,4.7 mas\,yr$^{-1}$ for S\,Ori\,73) are larger than the motion of $\sigma$\,Orionis, rendering S\,Ori\,70 and 73 cluster membership uncertain. From our survey, we identified one new photometric candidate with $J$\,=\,21.69\,$\pm$\,0.12 mag and methane color consistent with spectral type $\geq$\,T8.}
{S\,Ori\,73 has colors similar to those of T3--T5 field dwarfs, which in addition to its high proper motion suggests that it is probably a field dwarf located at 170--200 pc. The origin of S\,Ori\,70 remains unclear: it can be a field, foreground mid- to late-T free-floating dwarf with peculiar colors, or an orphan planet ejected through strong dynamical interactions from $\sigma$ Orionis or from a nearby star-forming region in Orion.}
\maketitle

\section{Introduction}
The discovery of isolated planetary mass objects \citep{lucas00,zapatero00,zapatero02sori70}, i.e. free-floating substellar bodies with
masses below the deuterium bur\-ning mass limit (M\,$\sim$\,12\,M$_{\rm Jup}$, \citet{saumon96,spiegel11}), challenges our understanding of the substellar formation processes. The proper characterization of young planetary-mass objects and the analysis of their cluster membership in known star-forming regions are necessary for an accurate derivation of the substellar mass function. The $\sigma$ Orionis cluster, with an age around 3 Myr \citep{sherry08}, low extinction \citep{bejar04a}, solar metallicity \citep{hernandez08}, and located at a distance of 352 pc \citep{perryman97}, offers an excellent ground to carry out these stu\-dies \citep{bejar01}. Previous photometric searches have revealed the presence of two T-type member candidates, S\,Ori\,70 (S\,Ori\,J053810.1-023626, \citet{zapatero02sori70}) and S\,Ori\,73 (S\,Ori\,J053814.5-024512, \citet{bihain09}). The former object has a spectral type T5.5$\pm$1 measured from its near-infrared spectrum co\-ve\-ring the $H$- and $K$-bands. Recent studies \citep{scholz08,zapatero08} indicates that S\,Ori\,70 has $J -K_s$ and {\sl Spitzer} colors redder than the typical colors of field T dwarfs, suggesting a low-gravity atmosphere. On the contrary, the neutral $J-K_s$  and the mid-infrared colors of S\,Ori\,73 do not appear to deviate significantly from the field T dwarf sequence. This object awaits confirmation of the presence of methane absorption in its atmosphere. Given its low brightness (S\,Ori\,73 is about 1 mag fainter than S\,Ori\,70 in the near-infrared wavelengths), spectroscopy would demand long integrations with large-size telescopes. Methane imaging \citep{rosenthal96,herbst99,tinney05} is a reasonable alternative method to deep into the at\-mos\-phe\-ric properties of S\,Ori\,73. The cluster membership confirmation for both S\,Ori\,70 and 73 is relevant to assessing the form of the $\sigma$\,Orionis mass function at planetary masses, since if both sources are true cluster members their masses are estimated to be in the interval 3--7 M$_{\rm Jup}$. Additional T-type sources are identified in other s\-te\-llar clusters and associations of different ages, like the Pleiades \citep{casewell07,casewell10}, the Trapezium \citep{lucas06}, IC\,348 \citep{burgess09}, and $\rho$ Oph \citep{marsh10,alvesdeoliveira10}.

Here we report on deep near-infrared and optical images co\-llec\-ted with 8--10-m class telescopes that will allow us to carry out a new ``pencil" search for additional T-type cluster member candidates of $\sigma$\,Orionis, to study the nature of S\,Ori\,73, and to derive accurate proper motions of the two currently known T dwarf cluster member candidates. Observations and data reduction are presented in section~2, section~3 presents the photometric and astrometric characterization of S\,Ori\,70 and 73 along with their cluster membership assessment, and the new search is des\-cri\-bed in section~4. Finally, we summarize the conclusions and final remarks in section~5.

\section{Observations}

\begin{table*}[!ht]
\caption[]{Log of optical and near infrared observations.}
\label{log}
$$
\begin{array}{p{0.9cm}p{1.5cm}p{2.8cm}p{1.2cm}p{0.85cm}p{3.8cm}p{1.5cm}p{1.5cm}p{1.5cm}p{0.9cm}}
\hline
\noalign{\smallskip}
Tel. & Instrument & $~~~~~~\alpha$ (J2000)~~$\delta$ & Field & Filter & Observing Date & Exposure & Seeing & Compl. & Limit \\
&&($^{\rm h~~~m~~~s}$)  ~~~~($^{\rm o~~~'~~~''}$)&&&&(s)&(arcsec)& (mag) & (mag) \\
\noalign{\smallskip}
\hline
\noalign{\smallskip}
\hline
\noalign{\smallskip}
GTC  &  OSIRIS & 05 38 26.8 $-$02 46 42 & S\,Ori\,73 & $i'$ & 2009 Oct 16, 2010 Jan 11 & 12636  & 0.95 & 25.0 & 26.0\\
          &  OSIRIS & 05 38 17.9 $-$02 37 08  & S\,Ori\,70 & $i'$  & 2009 Oct 13--15, 2009 Nov 19 & 12506 & 0.90 &25.0 & 26.0\\
          \noalign{\smallskip}
 \hline
 \noalign{\smallskip}
VISTA & VIRCAM & 05 38 20.6 $-$02 47 46 & $\sigma$\,Ori &$Z$ & 2009 Oct 20, 21 & 6084 & 0.80 & 22.6 & 23.2\\
 & & & & $Y$ & 2009 Oct 20 & 1008 &0.90 & 21.0 & 21.4\\
 & & & & $J$ & 2009 Oct 19, 20 & 2112 & 0.90 & 21.4 & 21.8\\
 & & & & $H$ & 2009 Oct 20 & 288 &0.90 & 19.6 & 20.0\\
 & & & & $K_s$ & 2009 Oct 20 & 288 & 0.70 & 18.6 & 19.1\\
 \noalign{\smallskip}
 \hline
 \noalign{\smallskip}
VLT & HAWK-I  & 05 38 08.5 $-$02 47 12 & S\,Ori\,73 &$J$ & 2008 Sep 19 & 160 & 0.64 & 22.5 & 23.2\\
 & & & & $H$ & 2008 Dec 8 & 8410 & 0.34 & 22.8 & 23.5\\
 & & & & $CH_{\rm 4off}$ & 2009 Feb 24, 2009 Mar 28 & 27000 & 0.51 & 22.2 & 23.0\\
VLT & HAWK-I & 05 38 03.3 $-$02 35 07 & S\,Ori\,70 &$J$ & 2008 Oct 27 & 160 & 0.75 & 21.7 & 22.2\\
 & & & & $H$ & 2009 Mar 24 & 8410 & 0.52 & 21.5 & 22.2\\
 & & & & $CH_{\rm 4off}$ & 2009 Mar 16 & 13500 & 0.57 & 21.0 & 21.8\\
VLT & HAWK-I & 05 38 41.5 $-$02 56 57& & $J$ & 2009 Feb 11 & 160 & 0.63 & 22.0 & 22.6\\
 & & & & $H$ & 2009 Mar 30 & 8410 & 0.75 & 21.0 & 21.5\\
 & & & & $CH_{\rm 4off}$ & 2009 Mar 29 & 12615 & 0.65 & 20.5 & 21.0\\
\noalign{\smallskip}
\hline
\end{array}
$$
\end{table*}

\subsection{Near-infrared data \label{obs}}

\begin{figure}
\centering
\includegraphics[width=0.47\textwidth]{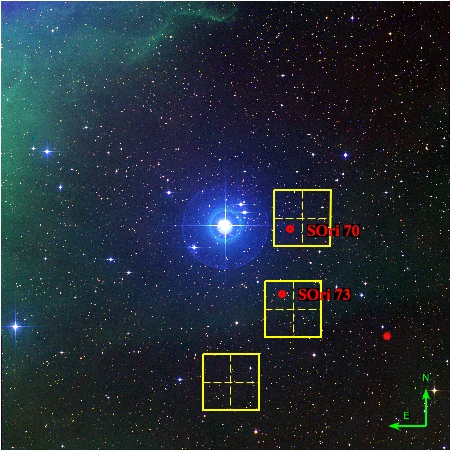}
\caption{False-color image of 1 deg$^2$ around the multiple, massive star $\sigma$\,Orionis. This image was produced by combining the optical $BVR$ POSS-II data. The three pointings of the HAWK-I survey are plotted as yellow squares, each square showing the positions of the four detectors of the HAWK-I instrument. The locations of the $\sigma$\,Orionis cluster T-type sources S\,Ori\,70 and 73 are indicated.}
\label{pointings_hawki}
\end{figure}

Near-infrared images were collected with the wide-field ca\-me\-ra HAWK-I \citep{hawki_1,hawki_2} installed at the Nasmyth A focus of unit 4 (Yepun) of the Very Large Telescope (VLT) on Cerro Paranal Observatory between September 2008 and March 2009. The on-sky field of view is 7.5 $\times$ 7.5 arcmin$^2$ per single shot covered by four Hawaii 2048\,$\times$\,2048 pixel detectors, which are separated by a cross-shaped gap of 15\arcsec. The instrument pixel scale is 0\farcs1064. Our HAWK-I data consist of three di\-ffe\-rent pointings, two of which are intended to i\-ma\-ge the known T-type cluster candidates S\,Ori\,70 and 73 (located to the west of the central, bright, and multiple star $\sigma$~Ori); the third poin\-ting is located about $20'$\,56\farcs8 to the south of the massive star. Figure~\ref{pointings_hawki} illustrates the three HAWK-I pointings. Images were taken using the broad-band filters $J$ and $H$, and the narrow-band filter $CH_{\rm 4off}$. The $CH_{\rm 4off}$ filter (central wavelength at 1.575~$\mu$m, width of 0.112~$\mu$m) covers wavelengths bluewards and out of the $H$-band strong methane absorption seen in T-type dwarfs \citep{burgasser02, geballe02}. A total of 4 ($J$), 29 ($H$), 30 ($CH_{\rm 4off}$, S\,Ori\,70), and 60 ($CH_{\rm 4off}$, S\,Ori\,73) individual images were acquired at different (random) telescope offsets within a box of 15\arcsec~in size for each pointing. This observing strategy allowed us to perform a proper Earth sky background contribution removal from the raw data. Images were obtained with typical seeing in the range 0\farcs34--0\farcs75; the field of S\,Ori\,73 was observed with the best seeing. We provide the log of the observations in Table~\ref{log}, where we include telescopes and instruments, central coordinates, field reference name (e.g., S\,Ori\,70), observing filter, observing dates, total exposure time, seeing, completeness and limiting magnitudes. 

\begin{figure}
\centering
\includegraphics[width=0.47\textwidth]{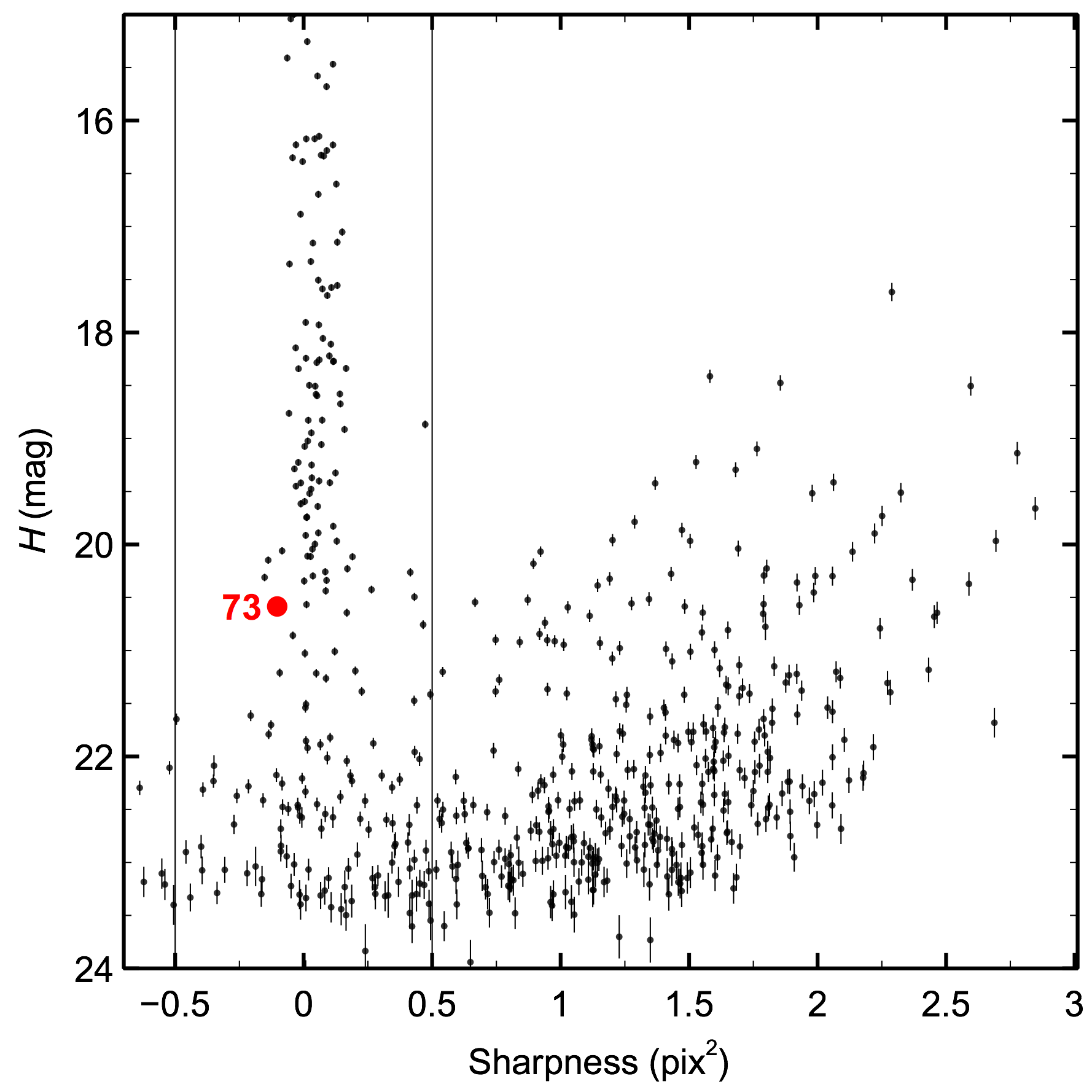}
\includegraphics[width=0.47\textwidth]{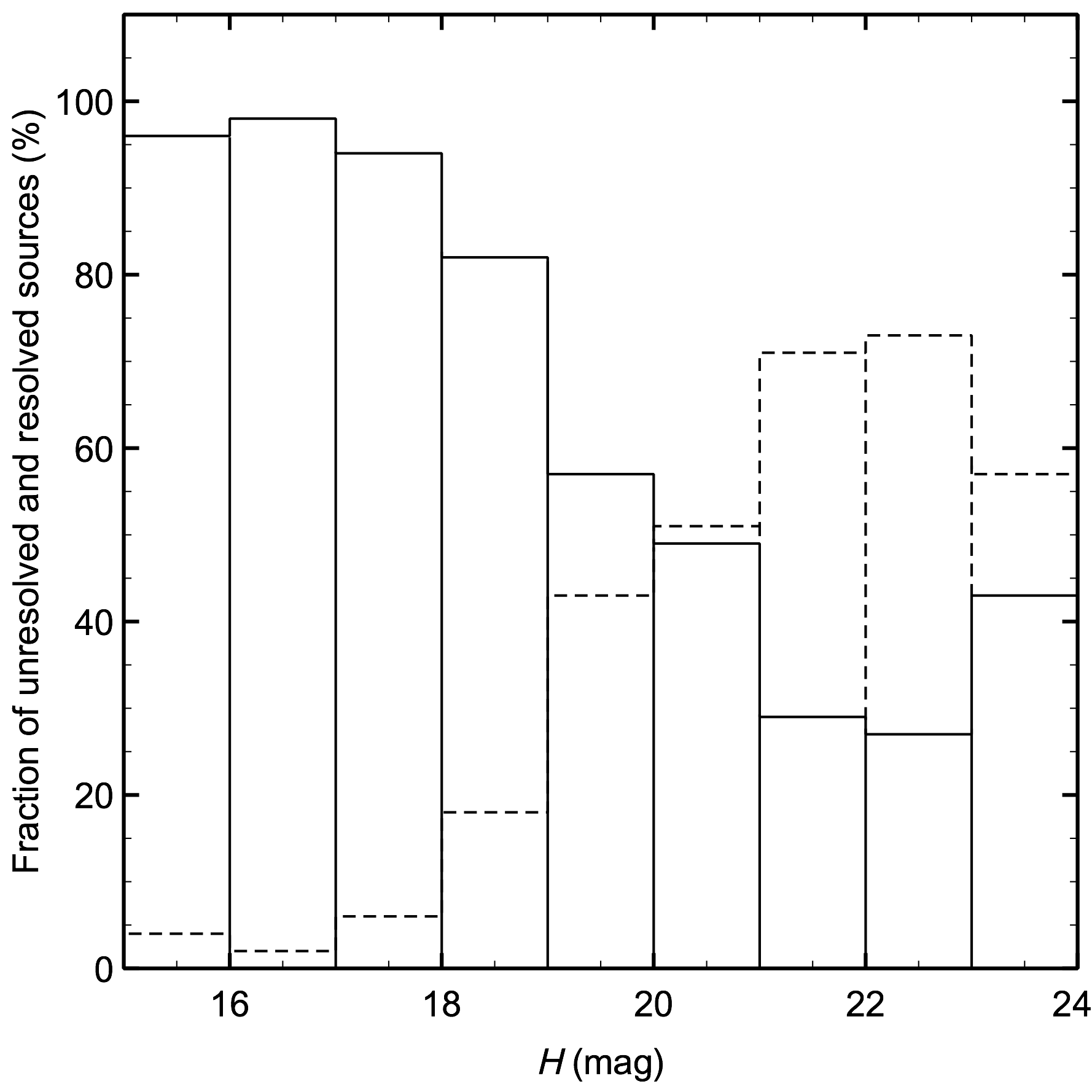}
\caption{Sharpness parameter (top panel) as a function of observed $H$-band magnitude for the HAWK-I S\,Ori\,73 pointing, S\,Ori\,73 plotted as a red filled circle. Our criterion to distinguish resolved from unresolved sources is indicated with vertical lines: object detections between the two vertical lines are supposed to be unresolved (e.g., point-like sources). The bottom panel illustrates the fraction of unresolved (solid line) and resolved (dashed line) sources per $H$-band magnitude bin.}
\label{sharp}
\end{figure}

HAWK-I data were reduced using standard techniques for the near-infrared within the {\sc iraf}\footnote{IRAF is distributed by National Optical Astronomy Observatories, which is operated by the Association of Universities for research in Astronomy, Inc., under contract to the National Science Foundation.} environment including dark, flat field, pixel mask correction and sky subtraction. Individual frames were aligned and stacked together to produce deep $J$, $H$, and $CH_{\rm 4off}$ images. By taking into account the gap between detectors and the offsets among individual images, we determined that the HAWK-I survey covers an area of 119.15~arcmin$^2$, 40.02~arcmin$^2$ of which was observed with a seeing of 0\farcs34 in the $H$ band.  The photometric analysis was performed by ob\-tai\-ning aperture and point spread function (PSF) photometry u\-sing routines of the {\sc daophot} package. We defined the stellar PSF of our images by fitting Gaussian functions to 10--20 single stars per detector and pointing. The good seeing quality of the data ($\le$0\farcs75) allowed us to discriminate between resolved and unresolved (point-like) sources using the ``sharpness'' parameter provided by the {\sc daophot} routines, which is defined as the difference between the square of the width of the object and the square of the width of PSF. Sharpness has values close to zero for single sources, large positive values for blended doubles and partially resolved galaxies and large negative values for cosmic rays and blemishes. Our adopted criterion flags an object detection with sharpness between $-0.5$ and $+0.5$ as a point-like source. As a precautionary measure, aimed at not losing any potential interesting source, we have also checked the photometry of object detections with sharpness values in the range 0.5--1.0 and unresolved (star-like) Gaussian full width at half maximum (FWHM), and the photometry of faint sources within a radius of 5--6\arcsec~from bright ($H$\,$\le$\,19\,mag) objects. Figure~\ref{sharp} displays the sharpness parameter for the S\,Ori\,73 HAWK-I pointing; the vertical dotted lines indicate our criterion to select unresolved sources. Also in Fig.~\ref{sharp}, we show the resulting fraction of resolved and unresolved object detections as a function of observed $H$-band magnitude. As expected, the number of extended (quite likely extragalactic) sources increases rapidly at faint magnitudes: at $H$$\sim$19~mag, about 50\%~of the objects appear resolved.

HAWK-I $JH$ instrumental magnitudes were converted into apparent magnitudes using the UKIRT Infrared Deep Sky Survey (UKIDSS)\footnote{The UKIDSS project is defined in \cite{lawrence07}. UKIDSS uses the UKIRT Wide Field Camera (WFCAM; \cite{casali07}) and a photometric system described in \cite{hewett06}. The pipeline processing and science archive are described in Irwin et al$.$ 2010, in prep. and \cite{hambly08}.} seventh data release (DR7). UKIDSS sources with errors smaller than 0.1 mag were cross-correlated against our data. The dispersion of the photometric calibration is 0.04 mag on average;  we have added this uncertainty quadratically  to the instrumental magnitude errors computed by {\sc iraf} for each source detection. For the $CH_{\rm 4off}$ filter, we obtained a photometric calibration relative to the $H$-band. Objects with $H$ magnitudes between 13 and 17 (probably Galactic stars of spectral types G--K according to their UKIDSS colors) were forced to have a null $H-CH_{\rm 4off}$ color. The typical dispersion of this procedure was 0.04\,--\,0.07 mag, which was added quadratically to the $H-CH_{\rm 4off}$ colors. In this way we will be able to re\-cog\-ni\-ze sources with an anomalous behavior in the $H$-band, i.e., sources with methane colors deviating from 0.0. We note that for a given HAWK-I detector, the zero points  are constant for the various pointings. However, the calibration shifts are slightly different (by up to 0.2 mag) from detector to detector.

We used additional near-infrared data in our study. As part of Visible and Infrared Survey Telescope for Astronomy (VISTA) science verification observations carried out between 2009 October 15 and November 02, VISTA \citep{vista_1,vista_2} has explored an area of $\sim$30 deg$^2$ in the Orion Belt using the broad-band $ZYJHK_s$ filters (Petr-Gotzens et al$.$ 2011, in prep.). This survey includes the $\sigma$\,Orionis cluster. VISTA is equipped with an infrared camera as dedicated ins\-tru\-ment with a pixel scale of 0\farcs339. In a single shot, VISTA provides on-sky coverage of 1.5 $\times$ 1.0 deg$^2$. All VISTA scien\-ce ve\-ri\-fi\-ca\-tion data were processed at the Cambridge Astronomy Survey Unit (CASU)\footnote{http://casu.ast.cam.ac.uk} using the VIRCAM pipeline (version 0.9.6). In our work we employed version 2.1 of the pho\-to\-me\-tric and astrometric VISTA Orion catalog, which provides magnitudes in all VISTA filters and right ascension and declination coordinates with a precision better than 0\farcs11. VISTA photometric and astrometric calibrations were performed by CASU using the 2MASS catalog \citep{skrutskie06}. Table~\ref{log} provides further instrumental and observing details of the VISTA Orion survey in the region of the $\sigma$~Orionis cluster.

Completeness and limiting magnitudes (see Table~\ref{log}) of the HAWK-I and VISTA surveys in $\sigma$~Orionis were estimated from the distribution of the instrumental magnitude error bars as a function of observed magnitudes (in a similar way as in \cite{bihain09}). We defined completeness magnitude as the magnitude bin at which the average instrumental photometric uncertainty is 0.1~mag, which corresponds to a signal-to-noise (S/N) ratio close to 10, and limiting magnitude is the magnitude bin with an average photometric error of 0.25~mag (S/N\,$\sim$\,4). We remark that the S\,Ori\,73 HAWK-I field, covering an area of 40.02 arcmin$^2$, was observed under very good conditions of seeing (0\farcs34) and atmospheric transparency leading to rather deep i\-ma\-ges, which are about 1 mag fainter than the remaining HAWK-I data (79.13 arcmin$^2$). We also note that the VISTA Orion survey provides longer exposure times at $Z$- and $J$-band for the $\sigma$~Ori region than for the rest of the Orion Belt area, therefore the resulting VISTA $\sigma$~Ori $Z$ and $J$ images are deeper.

\subsection{Optical data}

We obtained Sloan-$i'$ images of the fields of S\,Ori\,70 and 73 u\-sing the optical OSIRIS instrument \citep{cepa00} mounted on the Gran Telescopio de Canarias (GTC) at the Roque de los Muchachos Observatory between 2009 October and 2010 January. The ca\-me\-ra consists of a mosaic of two 4096\,$\times$\,4096 pix$^2$ detectors with a pixel projection of 0\farcs125 onto the sky; the OSIRIS field of view is 7.4\,$\times$\,7.4 arcmin$^2$ per single shot. A detector binning of 2\,$\times$\,2 pixels was used. We obtained a total of 52 images with an integration time of 60.5~s and 156 images of 60~s for the field of S\,Ori\,70; for the field of S\,Ori\,73, a total of 52 images of 61.5~s and 156 images of 60.5~s were acquired. The total on-source exposure times were thus 3.47 and 3.51~h for the fields of S\,Ori\,70 and 73, respectively.   Raw frames  were reduced using the {\sc ccdred} package within the {\sc iraf} environment. Individual frames were bias and dark current subtracted and flat-fielded. All frames were aligned, normalized at the same exposure time, and stacked together to produce deep $i'$-band images overlapping with the S\,Ori\,70 and 73 HAWK-I fields. The photometric analysis was performed u\-sing {\sc daophot} routines, which includes the selection of objects with point-like PSFs with the {\sc daofind} task (extended objects were mostly avoided) and aperture and PSF pho\-to\-me\-try. Instrumental magnitudes were transformed into observed magnitudes in the Sloan $i'$ system with observations of photometric standard stars from \cite{smith02} obtained during the same nights in photometric sky conditions. Raw seeing ranged from 0\farcs7 to 1\farcs1. The survey completeness and limiting magnitudes  (as defined in the previous section) are $i'$\,=\,25 and 26\,mag, respectively. Table~\ref{log} summarizes the log of the OSIRIS observations.


\section{Characterization of S\,Ori\,70 and S\,Ori\,73\label{7073phot}}
From our HAWK-I data, S\,Ori\,73 remains unresolved at a spatial resolution of 0\farcs34 (see Fig.~\ref{sharp}), thus suggesting that it is a likely Galactic source. OSIRIS, HAWK-I and VISTA pho\-to\-me\-try of S\,Ori\,70 and 73 is given in Table~\ref{photometry}. In VISTA, S\,Ori\,70 is detected only in $Y$ and $J$ with the $Y-J$ color expected for its T5.5 spectral type. Both VISTA and HAWK-I magnitudes in the $J$-band agree within the photometric errors, and they are also close to the values reported in the literature \citep{zapatero08}, suggesting that S\,Ori\,70's long-term photometric va\-ria\-bi\-li\-ty at around 1.2\,$\mu$m is below 0.15\,mag. 

On the contrary, the $J$-band observations of S\,Ori\,73 reported here are brighter by 0.14\,mag (HAWK-I) and 0.33\,mag (VISTA) than the value given by \cite{bihain09}. S\,Ori\,73 is detected only in the $J$ filter in the VISTA survey. Also the HAWK-I photometry in $H$-band of S\,Ori\,73 is 0.25\,mag brighter than the measurement provided in the discovery paper. These di\-ffe\-ren\-ces, slightly larger than the quoted photometric uncertainties ($\sim$0.1\,mag), hint at some variability for S\,Ori\,73. However, this requires confirmation using further data of higher accuracy.

\begin{table}[!ht]
\caption[]{Optical and near infrared photometry for S\,Ori\,70 and S\,Ori\,73.}
$$
\label{photometry}
\begin{array}{p{2cm}p{1.5cm}p{2cm}p{2cm}}
\hline
\noalign{\smallskip}
Instrument & Photometry  & S\,Ori\,70 & S\,Ori\,73\\
&&(mag)&(mag)\\
\noalign{\smallskip}
\hline
\noalign{\smallskip}
\hline
\noalign{\smallskip}
OSIRIS      & $i'$                 & $\geq$ 26.0 & $\geq$ 26.0 \\
VIRCAM    & $Y$                   & 20.85\,$\pm$\,0.13 &$\geq$ 21.47        \\
                 & $J$                    & 19.84\,$\pm$\,0.06 &20.58\,$\pm$\,0.10\\
HAWK-I     & $J$                   & 19.85\,$\pm$\,0.05 & 20.77\,$\pm$\,0.04 \\
                 & $H$                   & 20.10\,$\pm$\,0.05 &  20.58\,$\pm$\,0.05\\
                 & $H-CH_{\rm 4off}$& 0.61\,\,\,\,$\pm$\,0.07 & 0.28\,\,\,\,$\pm$\,0.07\\
\noalign{\smallskip}
\hline
\end{array}
$$
\end{table}

\begin{figure*}
\centering
\includegraphics[width=0.42\textwidth]{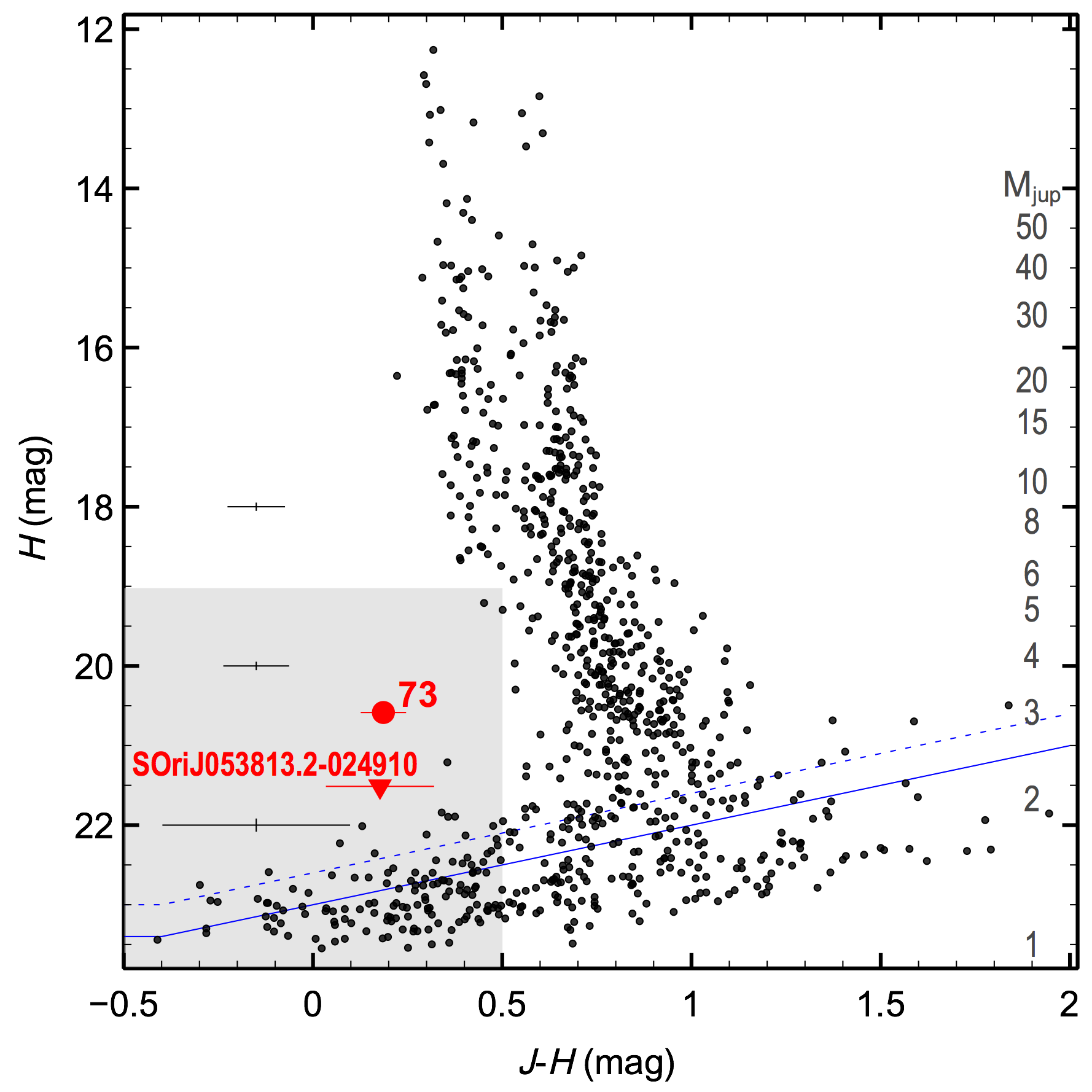}
\includegraphics[width=0.42\textwidth]{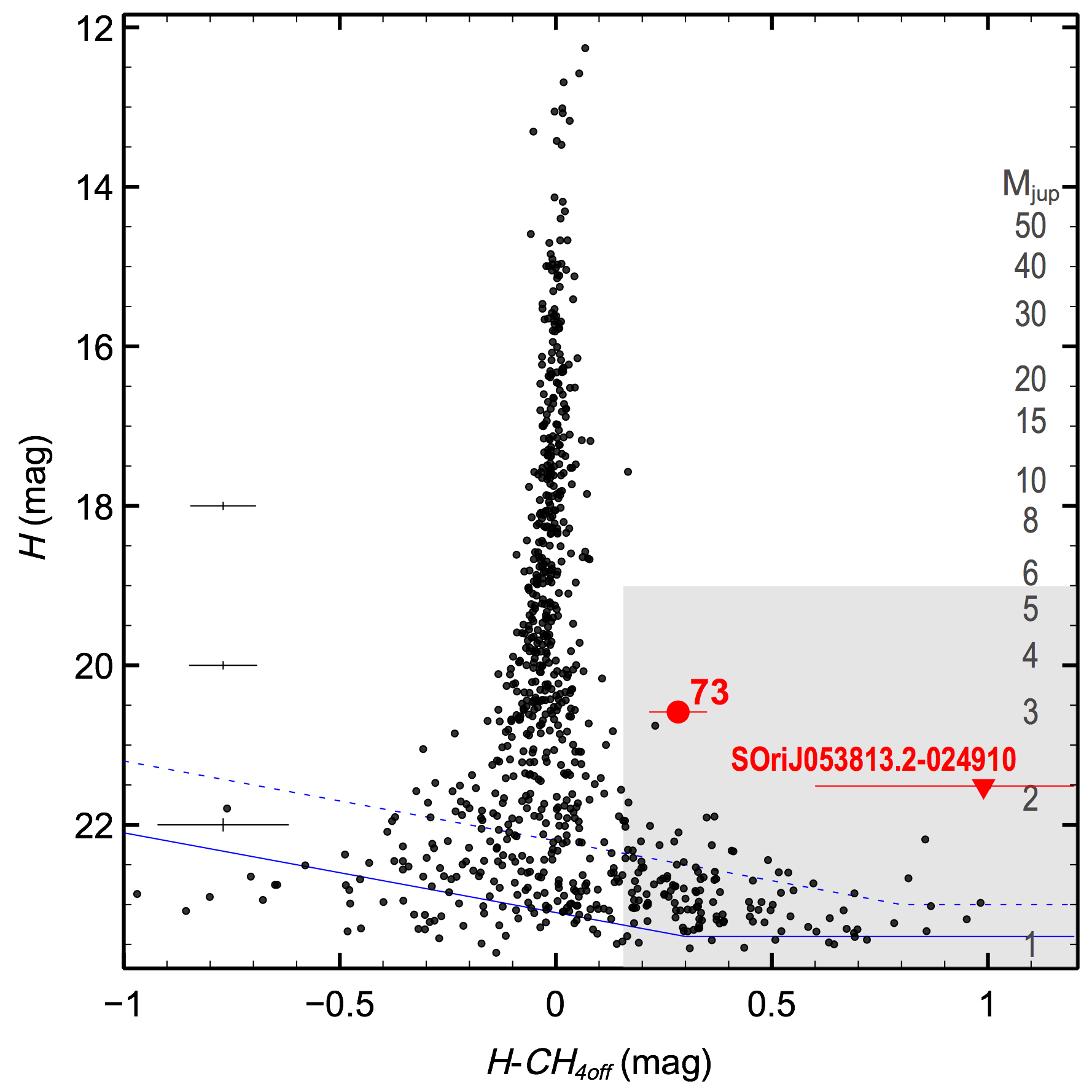}
\includegraphics[width=0.42\textwidth]{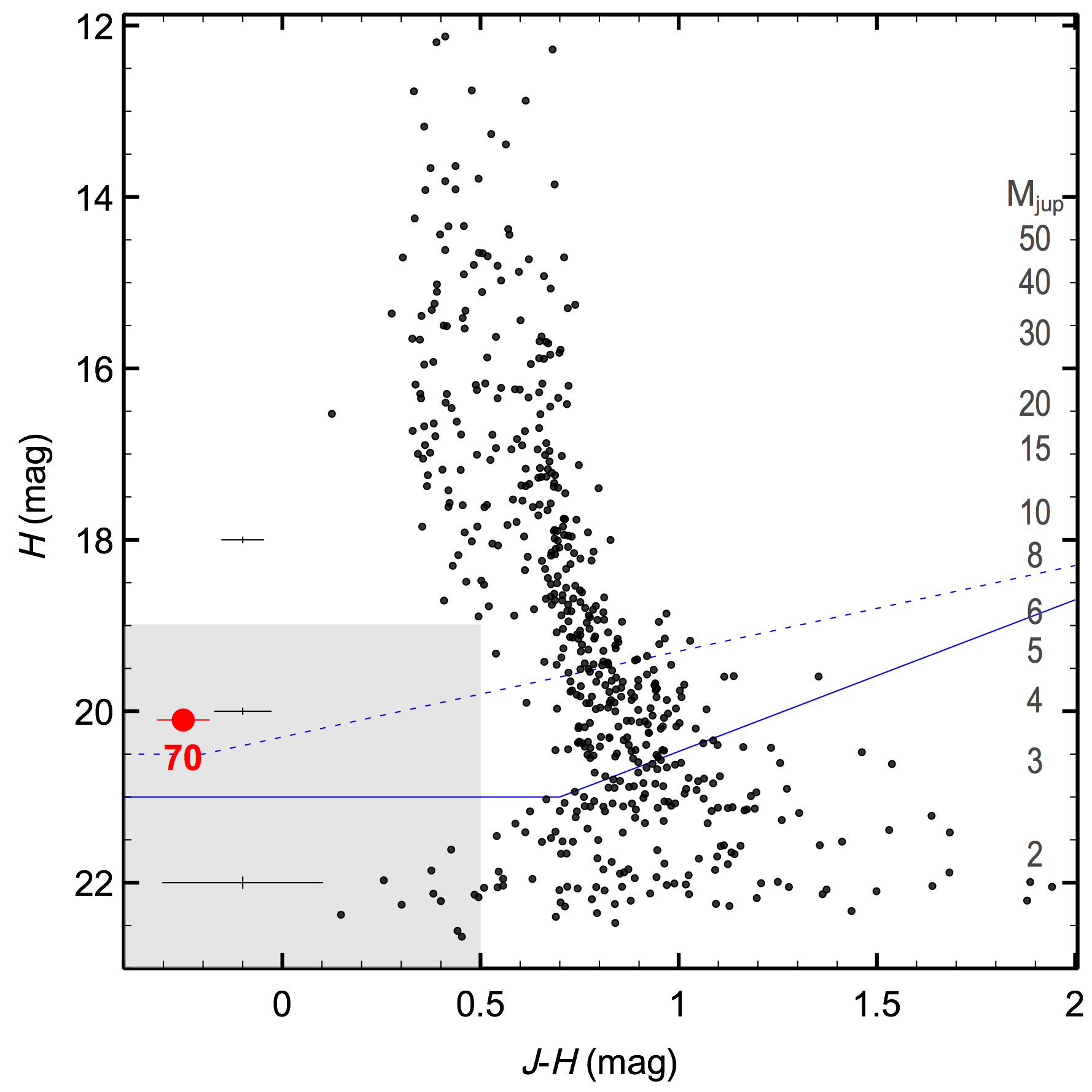}
\includegraphics[width=0.42\textwidth]{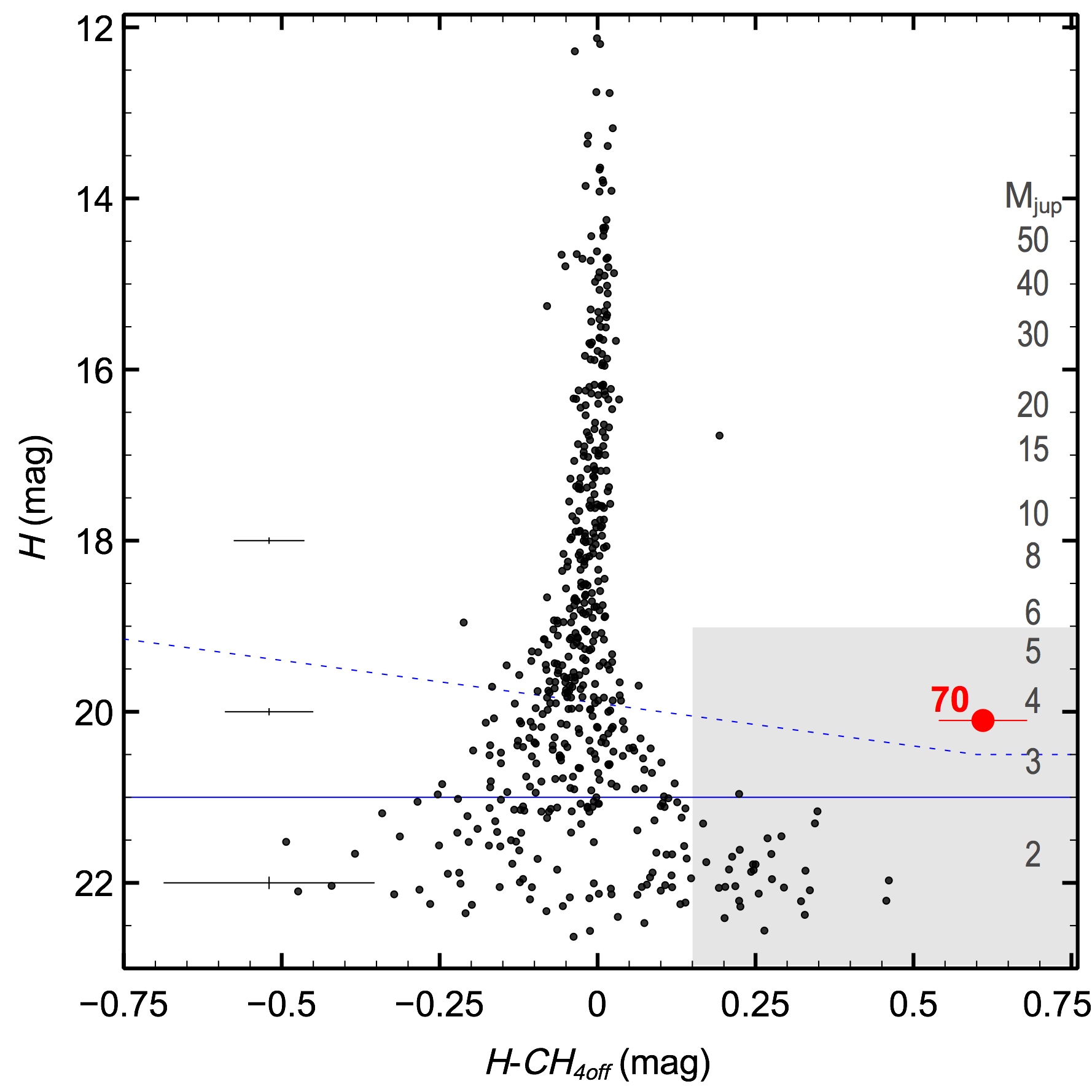}
\includegraphics[width=0.42\textwidth]{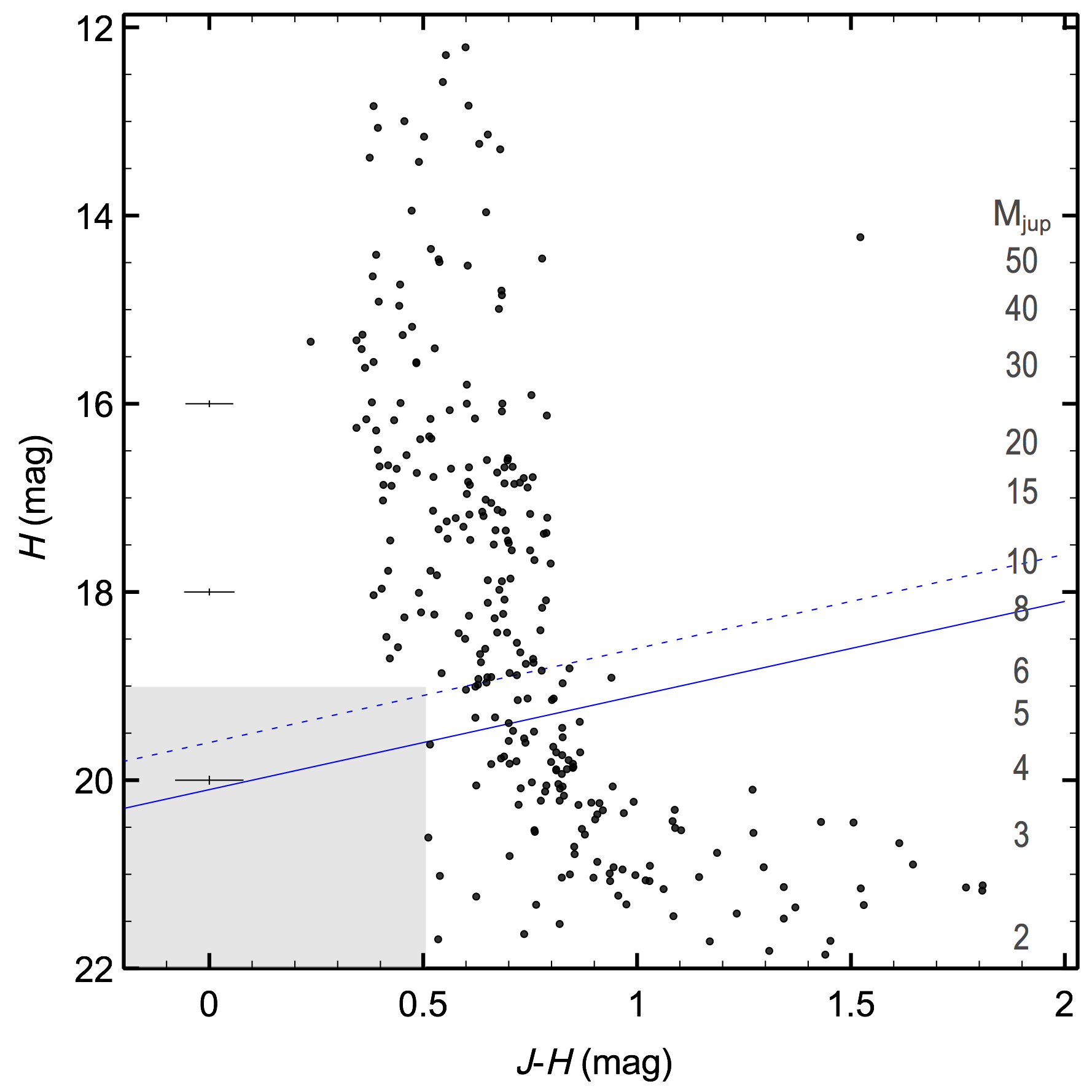}
\includegraphics[width=0.42\textwidth]{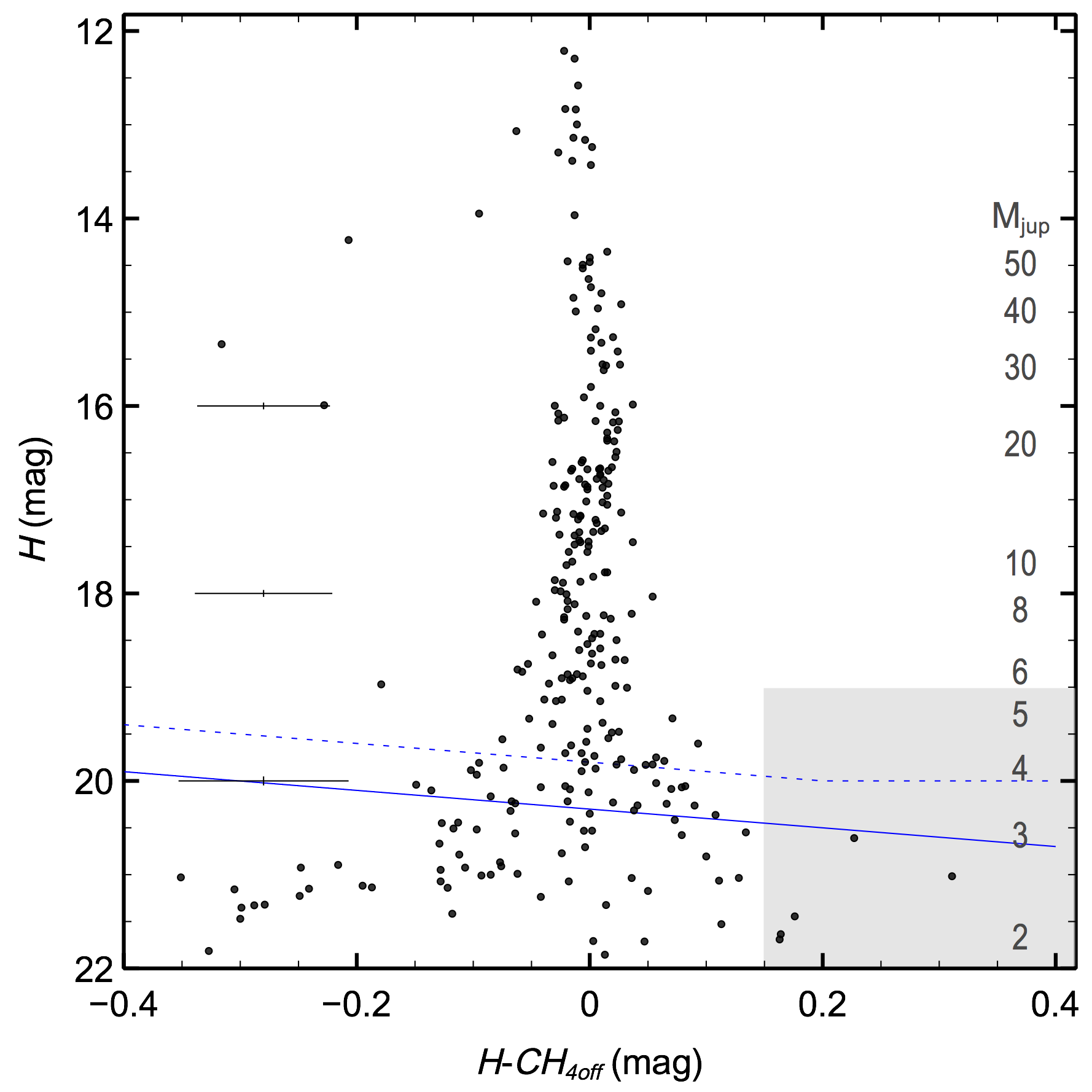}
\caption{Color-magnitude diagrams resulting from the HAWK-I survey. The top panels correspond to the S\,Ori\,73 pointing, middle panels display the field around S\,Ori\,70, and the lower panels illustrate the photometry of the third pointing. Sources discussed in this work are plotted as red circles (S\,Ori\,70 and 73) and with a red triangle (new T-type candidate). Completeness and limiting magnitudes are indicated by a dashed and a solid line, respectively. Our photometric criteria to select new methane candidates are shown with a gray-shaded region. Average error bars are given to the left, and masses (M$_{\rm Jup}$) predicted for the age of 3 Myr are labelled on the right side of the diagrams. }
\label{cm}
\end{figure*}

Figure~\ref{cm} displays the resulting color-magnitude diagrams of our HAWK-I data. Only unresolved sources are shown for the clarity of the plots. The location of objects, such as  S\,Ori\,70 and 73 that will be discussed in the following sections is marked with different symbols and colors. We use Fig.~\ref{cm} to study whether S\,Ori\,73 has methane absorption in the $H$-band and to search for new $\sigma$\,Ori T-type candidates in the cluster area of 119.15 arcmin$^2$ covered by HAWK-I. This area is twice as large as the one explored by \cite{zapatero02sori70}, and seven times smaller than the area surveyed by \cite{bihain09}, two previous deep searches for T dwarfs in the cluster that yielded the findings of S\,Ori\,70 and 73, respectively. 

\subsection{Methane absorption}
One of our objectives is to confirm the presence of methane absorption in the atmosphere of the $\sigma$~Ori T-type pho\-to\-me\-tric candidate S\,Ori\,73, and to constrain its spectral type. Given its low brightness, spectroscopy (even at near-infrared wavelengths) demands quite long integration times with current facilities. \cite{tinney05} and \cite{goldman10} have recently shown that methane imaging can be very useful to cha\-rac\-te\-ri\-ze T dwarfs since the methane color strongly depends on the T spectral subtypes (e.g., see Fig.~4 of \cite{goldman10}). From Fig.~\ref{cm}, it becomes apparent that S\,Ori\,70 and 73 are outliers. They have $H-CH_{\rm 4off}$ colors redder than point-like field sources of similar brightness, which we interpret as due to the presence of methane absorption in their atmospheres. These data thus confirm the ``methane'' nature of S\,Ori\,73. The $H-CH_{\rm 4off}$ color is expected to become redder for stronger methane absorption (i.e., in\-crea\-sing spectral types). S\,Ori\,73 has a $H-CH_{\rm 4off}$ index bluer than S\,Ori\,70 indicating less methane absorption intensity, i.e. earlier spectral type. 

To estimate the spectral typing of S\,Ori\,73 we used our HAWK-I data and a procedure similar to the one described by \cite{goldman10}. The $H-CH_{\rm 4off}$ color from HAWK-I  needs to be calibrated against spectral type. For this, we retrieved near-infrared spectra of main-sequence stars and brown dwarfs of F- through L-types from the archives available in the literature \citep{rayner09,cushing05}\footnote{http://irtfweb.ifa.hawaii.edu/spex/IRTFSpectralLibrary/index.htm}. Most of these spectra were taken with the medium-resolution spectrograph SpeX at the NASA Infrared Telescope Facility (IRTF) on Mauna Kea; they cover the spectral range of interest to us ($\sim$1.3--2.0\,$\mu$m) and have a resolving power of $R\sim$\,2000 and good S/N. To complete the spectral coverage towards cooler temperatures, spectra of field T-type dwarfs were retrieved from the libraries by \cite{knapp04, chiu06, golimowski04}\footnote{http://staff.gemini.edu/$\sim$sleggett/LTdata.html}. These data, obtained with SpeX and the UKIRT Imaging Spectrometer and Cooled Grating Spectrometer, have a lower spectral resolution of $R\sim$\,150. We use spectral types measured from optical data for the F--L sources, and from near-infrared data based on the classification scheme of \cite{burgasser02} for the T objects. All these spectra were conveniently convolved with the HAWK-I $H$ and $CH_{\rm 4off}$ filter transmission profiles and were flux integrated to derive the $H-CH_{\rm 4off}$ indices. The average color of the G--K stars ($J-H$\,$ \sim$\,0.4--0.6 mag) was artificially set to 0.0 to keep the same photometric reference system as for the HAWK-I data (see section~\ref{obs}). We have estimated that the uncertainty of this procedure is $\pm$\,0.1~mag in the methane color. We did not use the spectral library of T dwarfs by \cite{mclean03} because the integration of the spectra systematically yielded very red methane colors that could not be reconciled with the values obtained from the other spectral libraries. 

\begin{figure}[ht!]
\centering
\includegraphics[width=0.47\textwidth]{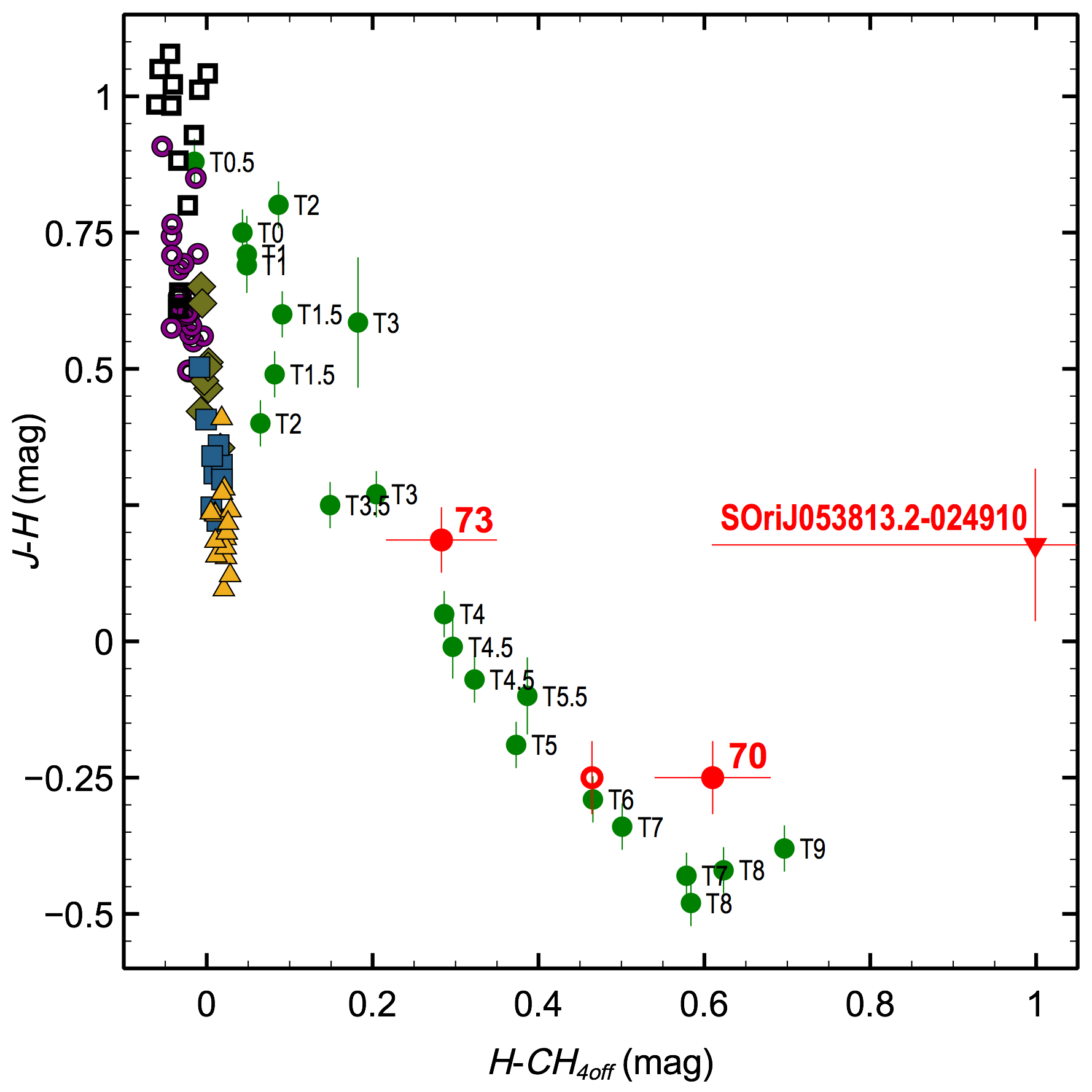}
\caption{Color-color diagram (including the HAWK-I methane color) for T (green), L (black), M (magenta), K (olive green), G (blue) and F (ye\-llow) type field sources. The photometry of S\,Ori\,70 and 73 is plotted as red filled circles and the new T-type candidate is shown with a red filled triangle. The red open circle corresponds to the synthetic methane color of S\,Ori\,70 derived from the Keck spectrum published in the literature (see text).}
\label{cc}
\end{figure}

The resulting diagram displaying the $H-CH_{\rm 4off}$ values of field F--T sources as a function of the $J-H$ index is shown in Fig.~\ref{cc}. As expected, T (``methane'') dwarfs are clearly recognized by their position in the diagram. Whereas stars and L dwarfs have $H-CH_{\rm 4off}$ colors between $-0.1$ and $\sim$0.03~mag, sub\-ste\-llar objects with types $\ge$T3 are characterized by $H-CH_{\rm 4off} \ge$ 0.1~mag and $J-H$ typically below 0.5~mag. The location of S\,Ori\,70 and 73 is indicated in Fig.~\ref{cc}. The methane color of S\,Ori\,73 is consistent with T4 spectral type, for which we estimate an uncertainty of one subtype. The photometric methane color of S\,Ori\,70 resembles that of T8\,$\pm$\,1 field dwarfs. This spectral classification is marginally consistent and cooler than the one (T5.5\,$\pm$\,1) from the literature \citep{zapatero02sori70}. We also integrated the Keck spectrum of S\,Ori\,70 \citep{zapatero02sori70} to derive its synthetic methane color; it is shown as an open circle in Fig.\ref{cc}. This computed color yields a classification of T6 (in agreement with the spectroscopic measurement); its a\-sso\-cia\-ted error bar is rather large since the signal-to-noise ratio of the spectrum is poor. Both purely photometric and ``integrated" measurements are consistent within the quoted uncertainties.

\subsection{Proper motions}
The HAWK-I and VISTA data add to the increasing battery of near-infrared images containing S\,Ori\,70 and 73. The first i\-ma\-ges (including the discovery frames) available for these two T-type sources were obtained several years (3.4 to $\sim$8 yr) before HAWK-I and VISTA observing runs, thus providing sufficient time baseline to carry out accurate proper motion measurements. The HAWK-I $H$-band and VISTA $J$-band images act as the second epoch data in our proper motion analysis. Regarding S\,Ori\,70, the first epoch frames are those presented in \citet{zapatero02sori70,zapatero08}: the $K_s$ image obtained with the NIRC instrument on the Keck~I telescope (2001 December 29, pixel size of 0\farcs15), and the $J$-band frame obtained with the Omega-2000 ($\Omega$2000) instrument on the 3.5-m telescope of the Calar Alto observatory (2005 October 22, pixel size of 0\farcs45). The first epoch $J$-band image of S\,Ori\,73 was taken with the ISAAC ins\-tru\-ment on the VLT (2001 December 10, pixel size of 0\farcs15) presented in \citet{caballero07} and re-reduced in \citet{bihain09}. The time se\-pa\-ra\-tion between first and second epoch data is listed in Table~\ref{pm} for each possible pair. Data of all epochs have good seeing qua\-li\-ty, typically below 1\arcsec, and both targets were detected with S/N greater than 10 in all frames.

\begin{figure*}[ht!]
\centering
\includegraphics[width=0.42\textwidth]{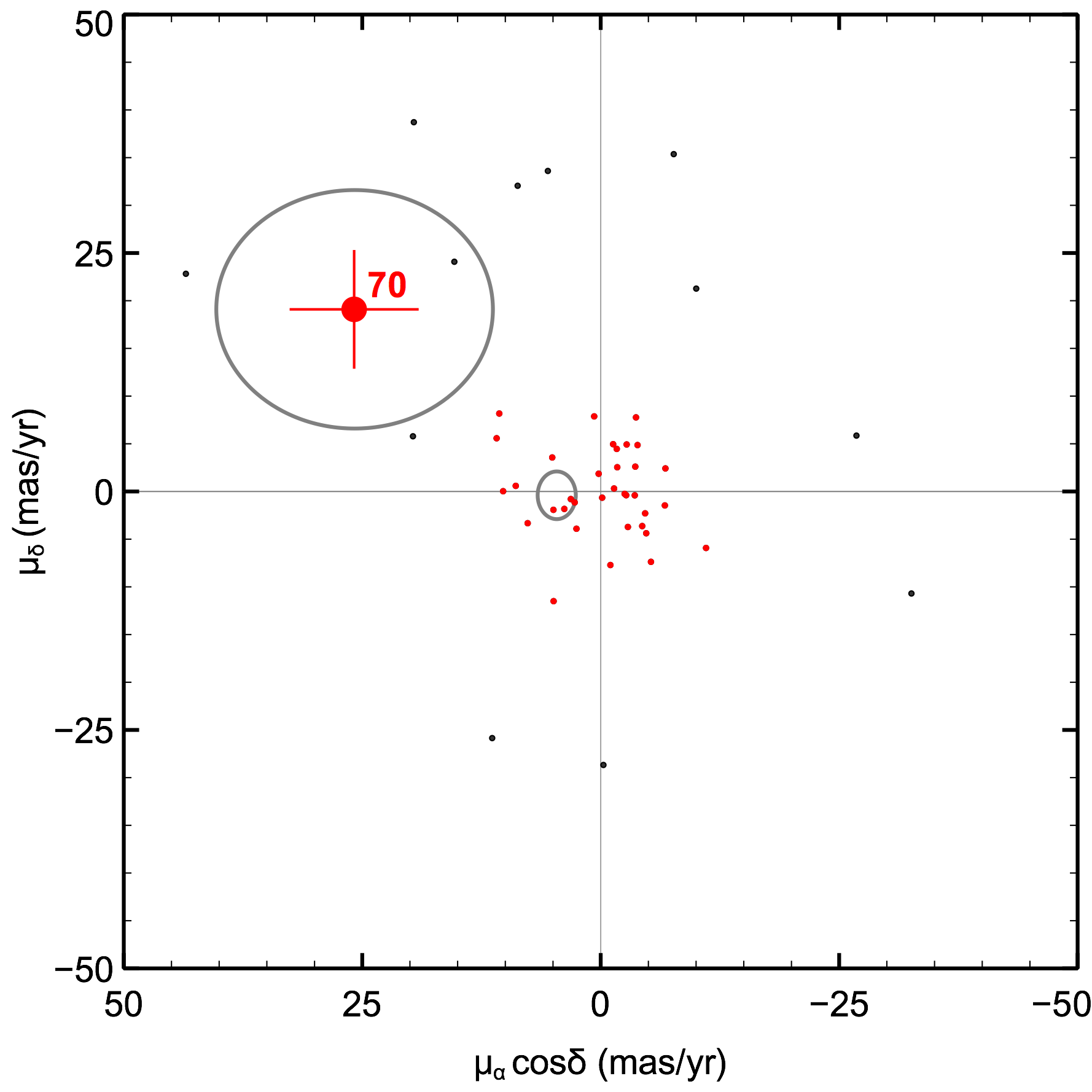}
\includegraphics[width=0.42\textwidth]{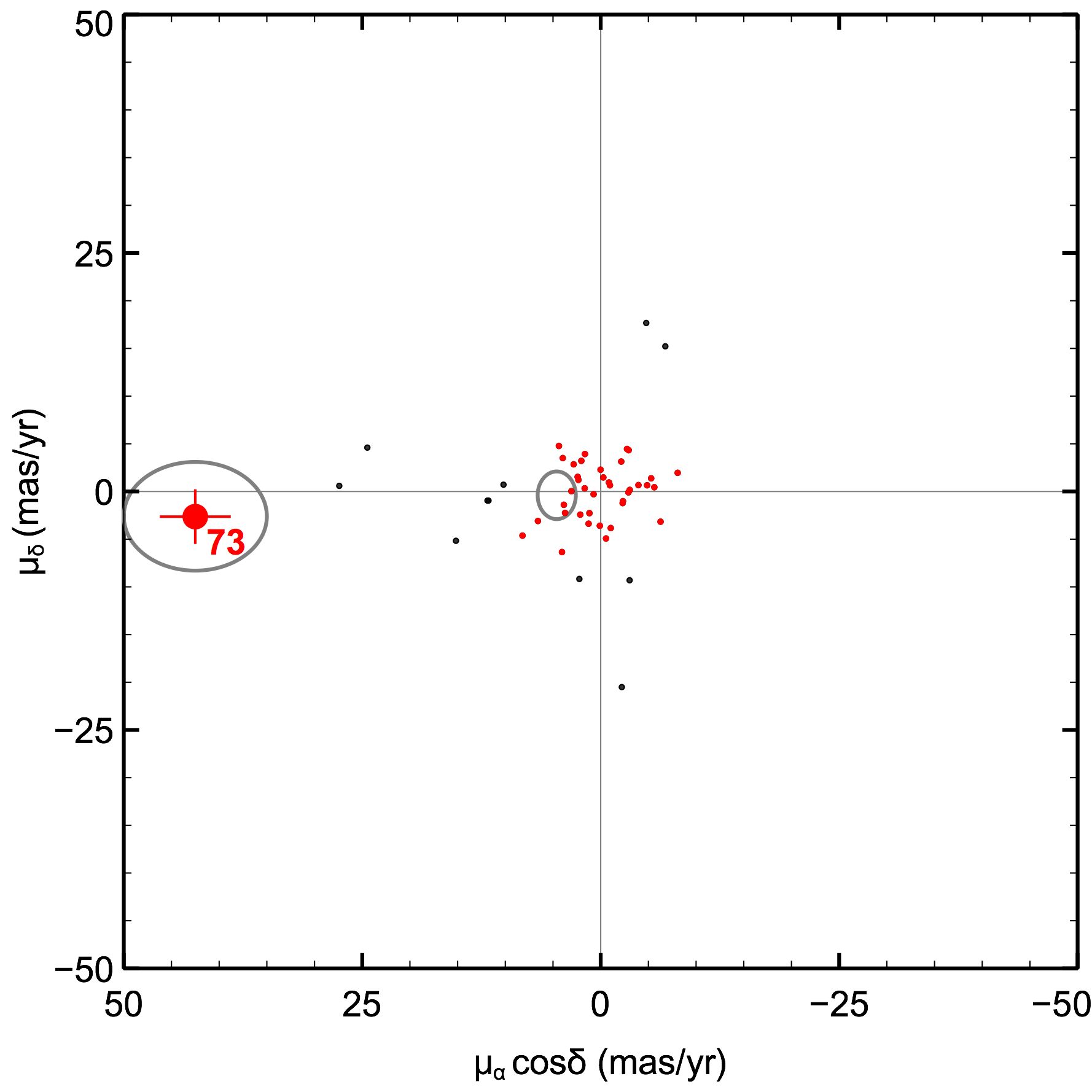}
\caption{Proper motion diagrams for S\,Ori\,70 (left panel, $\Omega$2000/HAWK-I data) and S\,Ori\,73 (right panel, ISAAC/HAWK-I data). The two objects of interest are labelled. All identified sources within an area of 4 arcmin$^{2}$ around the targets are plotted as small dots. The sources used as astrometric references are shown with red tiny dots. The ellipses around S\,Ori\,70 and 73 represent the 2-$\sigma$ proper motion uncertainties. The {\sl Hipparcos} motion of the $\sigma$\,Orionis cluster is depicted with a solid-line ellipse. }
\label{pmfig}
\end{figure*}

We determined the proper motions of S\,Ori\,70 and 73 by comparing their $(x,y)$ pixel positions with respect to several tens of unresolved sources within a projected distance of 2--3 arcmin from the targets. We checked that an area of 4--9 arcmin$^2$ provides a rea\-so\-na\-ble compromise between a high number (20--50) of reference sources (employed to establish the astrometric transformations) and re\-la\-ti\-ve\-ly ``small" field distortions (introduced by the instrument optics and detector pixel-to-pixel differences). The only exception is the NIRC instrument (S\,Ori\,70), which provides a rather small field of view (32\,$\times$\,32 arcsec$^2$) and a reduced number of reference sources (6). None of the reference sources is a known cluster member candidate, they are likely foreground and background objects. Pixel third-order polynomial transformations (second-order polynomials for NIRC) between epochs were calculated using the {\sc geomap} task within the {\sc iraf} environment; the dispersion of the transformations was typically below 0.3 pix (or 0\farcs06) for both the $x$ and $y$ axes after rejecting reference sources that deviate by more than 3$\sigma$ from null motion, where $\sigma$ denotes the dispersion of the astrometric transformations. The resulting pixel shifts were converted into proper motions ($\mu_\alpha\,{\rm cos}\,\delta$, $\mu_\delta$) by taking into account the time difference of the data, the pixel scale values, and the north-east orientation of the frames. Figure~\ref{pmfig} illustrates the results of the proper motion derivation for S\,Ori\,70 (Omega-2000 and HAWK-I data) and S\,Ori\,73 (ISAAC and HAWK-I data). We provide our proper motion measurements in Table~\ref{pm}. The proper motion uncertainties were calculated by adding qua\-dra\-ti\-ca\-lly the dispersion of the polynomial transformations and the errors of the targets centroids (0.03--0.12 pix). All individual determinations are consistent within the quoted uncertainties. In Table~\ref{pm} we also provide the mean proper motions of S\,Ori\,70 and 73 obtained by averaging individual measurements (excluding the measurement with the largest error bar for S\,Ori\,70). We will adopt these mean values in the next sections. Chromatic and differential refraction uncertainties are not contributing significantly to our proper motion measurements because we are using near-infrared data and polynomial astrometric transformations of degree two and higher. Following the equations and discussion of \citet{fritz10}, we estimate the chromatic and differential refraction errors to be below 1 mas.

\begin{table}[!ht]
\caption[]{Proper motion measurements for S\,Ori\,70 and S\,Ori\,73.}
\label{pm}
$$
\begin{array}{p{1.1cm}p{2.7cm}p{0.7cm}p{1.5cm}p{1.6cm}}
\hline
\noalign{\smallskip}
Object & Instruments & $\Delta t$&$\mu_\alpha\cos\delta$ & $\mu_\delta$\\
&&(yr)&(mas yr$^{-1}$) &(mas yr$^{-1}$) \\
\noalign{\smallskip}
\hline
\noalign{\smallskip}
\hline
\noalign{\smallskip}
S\,Ori\,70 & $\Omega$2000\,/\,HAWK-I &3.42    &25.8\,$\pm$\,6.8&19.1\,$\pm$\,6.3\\
                & $\Omega$2000\,/\,VIRCAM  &3.99  &25.1\,$\pm$\,17.8&8.1\,$\pm$\,11.9\\
                & NIRC\,\,/\,HAWK-I  &7.23  &25.5\,$\pm$\,5.2&27.1\,$\pm$\,5.0\\
                & NIRC\,/\,VIRCAM  &7.80  &28.8\,$\pm$\,6.2&17.7\,$\pm$\,6.9\\
                &Average			 &  &26.7\,$\pm$\,6.1&21.3\,$\pm$\,6.1\\
\hline
\noalign{\smallskip}
S\,Ori\,73 & ISAAC\,/\,HAWK-I  &6.99& 42.5\,$\pm$\,3.7&$-$2.6\,$\pm$\,2.9 \\
                & ISAAC\,/\,VIRCAM &7.85& 50.9\,$\pm$\,6.0& $-$10.0\,$\pm$\,6.5 \\
                & Average	&&46.7\,$\pm$\,4.9& $-$6.3\,$\pm$\,4.7 \\
\hline
\end{array}
$$
\end{table}

\subsection{Cluster membership\label{7073pm}}
Both S\,Ori\,70 and 73 present relatively distinct, high proper motions: $\mu\,=\,$34.1\,$\pm$\,8.6 mas\,yr$^{-1}$ and 47.1\,$\pm$\,6.8 mas\,yr$^{-1}$ with position angles of 51.4\,$\pm$\,14 deg and 97.7\,$\pm$\,7 deg, respectively (position angles are measured east of north). Our measurement of $\mu$ for S\,Ori\,70 is consistent at the level of 2.3-$\sigma$ with the proper motion value derived in \citet{zapatero08}. Here and from now on, $\sigma$ represents
the quadratic sum of the uncertainties associated to each proper motion component. The proper motion of S\,Ori\,70 and 73 appear to be larger than the proper motion of the central, multiple star $\sigma$\,Ori (\,$\mu\,=\,$4.7\,$\pm$\,1.0 mas\,yr$^{-1}$, position angle of 95\,$\pm$\,10 deg) as measured by Hipparcos \citep{perryman97} by factors of 4.5-$\sigma$ and 7.4-$\sigma$, respectively. These large differences are difficult to reconcile with the small (nearly null) motion of the $\sigma$\,Orionis cluster.

At the distance of $\sigma$\,Orionis (352 pc), the tangential ve\-lo\-ci\-ties ($v_{\rm tan}$) of S\,Ori\,70 and 73 would be 57\,$\pm$\,14.3 and 78.6\,$\pm$\,11.0 km\,s$^{-1}$, respectively. These values are larger than those of brighter cluster members. S\,Ori\,73 has a $v_{\rm tan}$ in excess of the typical $v_{\rm tan}$ values of field T dwarfs published by \citet{faherty09} while S\,Ori\,70 lies in the velocity range of known field dwarfs. If both sources were field objects, using their spectral classifications and observed $J$ magnitudes, they would be located at distances of 80--120 pc (S\,Ori\,70) and 170--200 pc (S\,Ori\,73), translating into $v_{\rm tan}$ values of 13--19.5 km\,s$^{-1}$ (S\,Ori\,70) and 38--45 km\,s$^{-1}$ (S\,Ori\,73). The $v_{\rm tan}$ of S\,Ori\,73 then becomes consistent with the tangential velocities of field T dwarfs \citep{faherty09}, which have ages of a few Gyr, and the $v_{\rm tan}$ of S\,Ori\,70 falls at the low values of the $v_{\rm tan}$ distribution of field T dwarfs.

\citet{jeffries06} and \citet{sacco08} showed that the radial velocity dispersion of $\sigma$\,Orionis cluster members with an average mass between 0.1 and 0.5 M$_\odot$ is 1.3 km\,s$^{-1}$, which translates into a cluster internal proper motion dispersion of 0.8 mas\,yr$^{-1}$ (1-$\sigma$) using 352\,pc as the cluster distance. This ve\-lo\-ci\-ty dispersion is a factor of $\sim$3 lower than the value found for the 1-Myr Orion Nebula cluster by \citet{furesz08}. \citet{bihain06} discussed that Pleiades substellar members with a mass around 30--50 times the mass of Jupiter (M$_{\rm Jup}$) have a proper motion dispersion of $\ge$\,4\,mas\,yr$^{-1}$, which the authors found consistent with the linear relationship of equipartition of e\-ner\-gy for a relaxed cluster. Given its total mass (between 150 and 250 M$_\odot$, \citep{hodapp09}), relatively low stellar density \citep{caballero08}, and young age ($\sim$\,3 Myr), the $\sigma$\,Orionis cluster has likely existed for a few crossing times and in consequence, the cluster may be dynamically unevolved. The expected re\-la\-xa\-tion time of $\sigma$\,Orionis is around 21--35 Myr,\footnote{In \citet{sherry04} the relaxation time of a cluster similar to $\sigma$ Orionis was estimated at 7 times the crossing time ($t_{\rm cross}$), where $t_{\rm cross}\leq 3-5$ Myr.} about ten times larger than the cluster age. By assuming the cluster is close to relaxation we estimate that the maximum proper motion dispersion expected for T-type members with a typical mass of 5 M$_{\rm Jup}$ is 5.5 mas\,yr$^{-1}$(1-$\sigma$) or 11.0 mas\,yr$^{-1}$ (2-$\sigma$). The motions of both S\,Ori\,70 and 73 appear to be beyond the conservative proper motion dispersion estimated for $\sigma$\,Orionis planetary-mass objects.

The full extension of the $\sigma$\,Orionis cluster is not properly determined yet, however various studies in the literature (e.g., \citet{sherry04,lodieu09a,bejar11}) have established that the majority of the cluster members ($\ge$80\%) are located at distances up to a radius of 0.5 deg from the central O-type multiple star. At the current velocities of S\,Ori\,70 and 73, these two objects would need only $\sim$1.0\,$\times$\,10$^5$ and $\sim$0.8\,$\times$\,10$^5$ yr to cross the entire cluster, which represents less than 10\%~of the cluster age. Possibly, the two sources have recently acquired their large motions through strong dynamical interactions with more massive members of the cluster or as a result of expulsions from multiple, interacting systems. In addition, the motions (and large $v_{\rm tan}$) of S\,Ori\,70 and 73 are not compatible with them being bound at large orbital separations from any cluster member in their surroundings, since their expected orbital velocities are necessarily smaller than the measured values. Their $v_{\rm tan}$ measurements also appear to be in excess of the cluster escape velocity at their position.

Because the $\sigma$\,Orionis cluster belongs to the Orion star-forming complex, occupying several tens of degrees on the sky, there is a possibility that either S\,Ori\,70 or 73 is an ejected object from a nearby star-forming region. Over the last 12 Myr, Orion has given birth to at least ten thousand stars contained in sub-groups and short-lived clusters (e.g., \citet{bally08}). We linearly traced back the motions of S\,Ori\,70 and 73 until they were beyond the limits of the Orion complex, and we searched for known star-forming regions that happen to lie along the projected motions. For S\,Ori\,73, no region was identified. However, the backward projected motion of S\,Ori\,70 ran at a separation of about ${23'}$ from the LDN\,1634 cometary molecular cloud about 6.0\,$\times$\,10$^5$ yr ago. This cloud, located west of the Orion Nebular Cloud, is probably shaped by the ultraviolet radiation from the Trapezium, and contains nine young stars (typically below 1 Myr) and three outflows, including the well-known HH 240/241 system \citet{bally09}. \citet{lee05} confirmed that LDN\,1634 is currently undergoing stellar formation.

The photometric and spectroscopic properties of S\,Ori\,70 and 73 can also give us additional information on their origins. The near- and mid-infrared colors of S\,Ori\,73 do not deviate from those of field T3--T5 dwarfs, which contrasts with the colors of S\,Ori\,70. As discussed in \citet{zapatero08}, S\,Ori\,70 displays red $J-K$ and {\sl Spitzer} colors qua\-li\-ta\-ti\-ve\-ly in agreement with the theoretical predictions for solar-metallicity, low-gravity atmospheres. The reddish nature of this object was confirmed by \citet{luhman08}, who found that it is slightly redder in $[3.6]-[4.5]$ than the reddest field dwarfs. \citet{scholz08} argued that this property may be attributed to the presence of a surrounding dusty disk. This would support the youth of S\,Ori\,70; we cannot rule out that it was formed in either $\sigma$\,Orionis or a nearby Orion star-forming region such as LDN\,1634. The colors of S\,Ori\,73 suggest that it has a high-gravity at\-mos\-phe\-re similar to field dwarfs of related spectral classification.

\begin{figure}[ht!]
\centering
\includegraphics[width=0.47\textwidth]{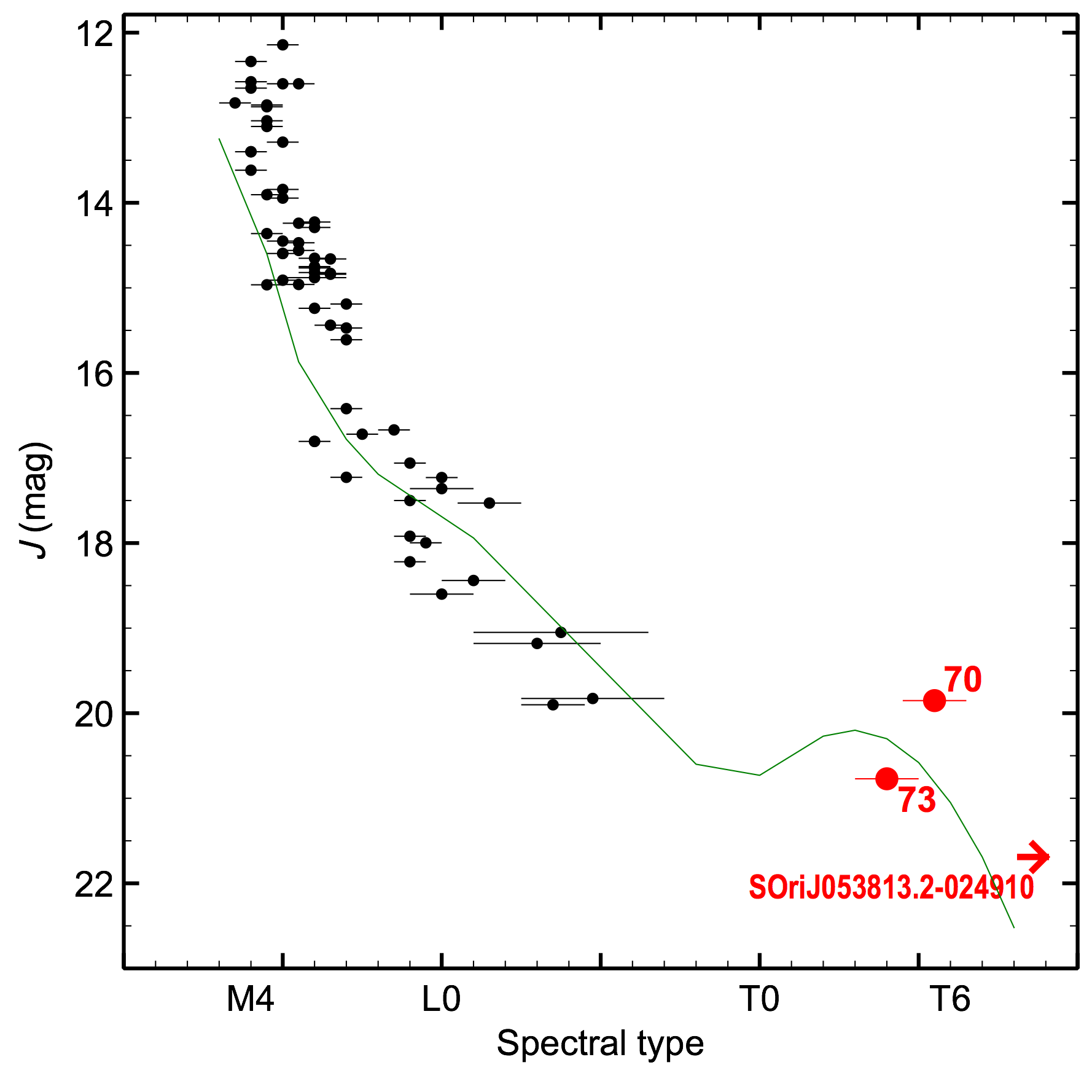}
\caption{Spectroscopic sequence of $\sigma$ Orionis cluster members (black dots). Data for mid-M to mid-L objects are taken from the literature (see \citet{zapatero08}). S\,Ori\,70, 73 and the new T candidate are labelled. The average sequence of field M-, L-, and T-type dwarfs is normalized at the late-M types of the cluster and plotted as a green solid line. We use it to ``predict" the location of $\sigma$ Orionis T-type members.}
\label{secuencia}
\end{figure}

The spectral sequence of $\sigma$\,Orionis is depicted in Fig.~\ref{secuencia}, where we include S\,Ori\,70 and 73. All spectral types were derived from observed low-resolution spectra available in the li\-te\-ra\-tu\-re, except for S\,Ori\,73 whose spectral type is obtained from methane photometry. Overplotted onto the data is the average sequence of field M-, L-, and T-dwarfs conveniently normalized at the late-M types of the cluster. We use the normalized field sequence to aid the interpretation of the locations of S\,Ori\,70 and 73 in Fig.~\ref{secuencia} with respect to the positions of warmer $\sigma$\,Orionis members. Despite of its cooler surface temperature, S\,Ori\,70 is 0.5--1 mag brighter than S\,Ori\,73 at all wavelengths studied, which is not an expected situation if both objects were true cluster members (in principle, considering single sources, fainter objects should be cooler at a given age and metallicity). However, from Fig.~\ref{secuencia} we cannot discard their membership in $\sigma$\,Orionis individually: S\,Ori\,70 and 73 appear slightly overluminous and underluminous than expected for T6- and T3-type members, res\-pec\-ti\-ve\-ly, yet their positions might be consistent with cluster membership if the $\sigma$\,Orionis sequence has an intrinsic photometric dispersion of about 1~mag at these faint magnitudes as suggested by the observed sequence delineated by M- and L-type members. This photometric dispersion may be caused by equal mass binaries, variability, slightly different ages, etc.

If neither S\,Ori\,70 nor 73 were members of $\sigma$\,Orionis but ``old" field dwarfs, we would derive a T4--T6 volume density of 1.44\,$\pm$\,1.00\,$\times$10$^{-2}$ dwarfs per pc$^{3}$ by taking into account the area of 790 arcmin$^2$ and the magnitude interval $J$\,=\,19\,--\,21.1 mag explored by \citet{bihain09}. This density is a factor of 2\,--\,10 higher than the average volume densities published in the li\-te\-ra\-tu\-re for similar spectral types \citep{reyle10,metchev08,burningham10}, suggesting that either S\,Ori\,70 or 73 is not a field object or that the density of field T dwarfs is higher in the direction towards the $\sigma$\,Orionis cluster.   If only one of the S\,Ori objects is a field dwarf, our volumen density determination for T4--T6 sources would decrease to values in better agreement with the literature, and the cluster mass function at planetary masses would look like the one depicted in Fig.~13 by \citep{bihain09}: the mass function appears to show a turnover below 6 M$_{\rm Jup}$. Cluster nonmembership of S\,Ori\,70 and 73 would have a dramatic impact in the cluster mass function: the mass bin corresponding to the interval 3.5--6 M$_{\rm Jup}$ would become ``empty", suggesting an abrupt end to the formation of free-floating substellar objects. This fact and the mass at which it happens are critical parameters for the theory of stellar and substellar formation processes. 

Based on all the  preceding discussion, we conclude that S\,Ori\,73 is probably not a member of the $\sigma$\,Orionis cluster; it seems to be a field T3--T5 dwarf at a distance of 170--200 pc in the line of sight towards Orion. The origin of S\,Ori\,70 remains unclear; it can be a field, foreground mid- to late-T free-floating dwarf with peculiar colors, or an orphan planet ejected from a nearby Orion star-forming region including $\sigma$ Orionis.


\begin{table*}[!ht]
\caption[]{Near infrared photometry for the new T-type candidate.}
\label{newobj}
$$
\begin{array}{p{3.2cm}p{1.8cm}p{1.8cm}p{1.8cm}p{1.5cm}}
\hline
\noalign{\smallskip}
Candidate & $J$ & $H$ & $J-H$ & $H-CH_{\rm4off}$ \\
&(mag)&(mag)&(mag)&(mag)\\
\noalign{\smallskip}
\hline
\noalign{\smallskip}
\hline
\noalign{\smallskip}
S\,Ori\,J053813.2$-$024910 &  21.69\,$\pm$\,0.12 & 21.51\,$\pm$\,0.07 & 0.18\,$\pm$\,0.13  & 0.99\,$\pm$\,0.39 \\
\hline
\noalign{\smallskip}
\end{array}
$$
\end{table*}

\begin{table*}[!ht]
\caption[]{Photometry of rejected candidates.}
\label{wrong}
$$
\begin{array}{p{3.2cm}p{1.8cm}p{1.8cm}p{1.8cm}p{1.8cm}p{1cm}}
\hline
\noalign{\smallskip}
Source & $J$ & $H$ & $J-H$ & $H-CH_{\rm4off}$& $i'-J$\\
&(mag)&(mag)&(mag)&(mag)&(mag)\\
\noalign{\smallskip}
\hline
\noalign{\smallskip}
\hline
\noalign{\smallskip}
S\,Ori\,J053803.6$-$024922 &  23.01\,$\pm$\,0.20 & 22.59\,$\pm$\,0.10 & 0.42\,$\pm$\,0.22  & 0.37\,$\pm$\,0.16 & $\leq$ 2.0\\ 
S\,Ori\,J053817.9$-$024631 & 22.80\,$\pm$\,0.18 & 22.32\,$\pm$\,0.06 & 0.48\,$\pm$\,0.19 & 0.40\,$\pm$\,0.10 & $\leq$ 2.2\\
S\,Ori\,J053810.0$-$024643 & 22.80\,$\pm$\,0.19 & 22.60\,$\pm$\,0.08 & 0.20\,$\pm$\,0.20 & 0.54\,$\pm$\,0.13 & $\leq$ 2.2\\
S\,Ori\,J053803.5$-$024542 & 22.74\,$\pm$\,0.18 & 22.25\,$\pm$\,0.08 & 0.49\,$\pm$\,0.20 & 0.36\,$\pm$\,0.13 & $\leq$ 2.3\\
S\,Ori\,J053810.1$-$024859 & 21.18\,$\pm$\,0.30 & 20.76\,$\pm$\,0.04 & 0.42\,$\pm$\,0.30 & 0.23\,$\pm$\,0.07 & $\leq$ 4.0\\
S\,Ori\,J053818.8$-$024516 & 22.86\,$\pm$\,0.19 & 22.98\,$\pm$\,0.11 & $-$0.12\,$\pm$\,0.22 & 0.98\,$\pm$\,0.14 & $\geq$ 3.1$^{\rm (a)}$\\
\hline
\end{array}
$$
\begin{list}{}{}
\item(a) This rejected candidate has a $J-K'$\,$\geq$\,0.5 mag.
\end{list}
\end{table*}

\section{New search for T-type cluster member candidates\label{busqueda}}

\begin{figure*}[ht!]
\centering
\includegraphics[width=0.42\textwidth]{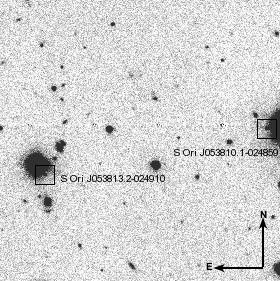} ~ \vspace*{0.1cm}
\includegraphics[width=0.42\textwidth]{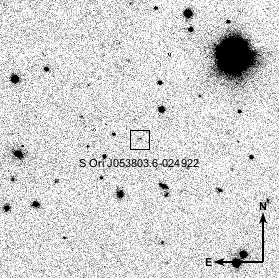}  \vspace*{0.1cm}
\includegraphics[width=0.42\textwidth]{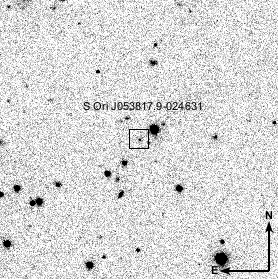} ~ \vspace*{0.1cm}
\includegraphics[width=0.42\textwidth]{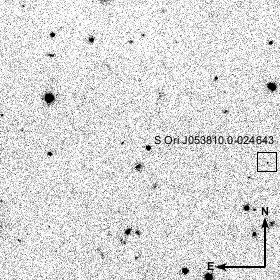} \vspace*{0.1cm}
\includegraphics[width=0.42\textwidth]{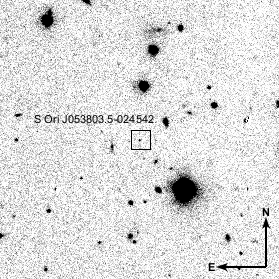} ~ \vspace*{0.1cm}
\includegraphics[width=0.42\textwidth]{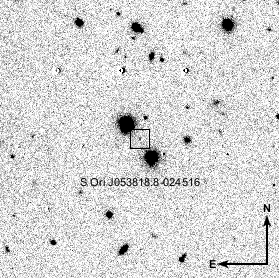} 
\caption{Finding charts (HAWK-I $H$-band, 1 $\times$ 1 arcmin$^2$ in size) for the new photometric T-type candidate (S\,Ori\,J053813.2$-$024910, upper left panel) and six rejected candidates.}
\label{charts}
\end{figure*}

To identify new unresolved sources with methane absorption from the HAWK-I color-magnitude diagrams illustrated in Fig.~\ref{cm}, we applied the following photometric criteria: $J$\,$\ge$\,19\,mag, neutral near-infrared colors, i.e., $J-H$\,$\le$\,0.5\,mag, and red methane index, i.e., $H-CH_{4\rm off}$\,$\ge$\,0.15\,mag. According to Fig.~\ref{cc}, the cut in the methane color allowed us to select sources with spectral types $\ge$\,T3. There are a total of 9 sources, in\-clu\-ding S\,Ori\,70 and 73, that fulfill our requirements. The 7 new objects are found in the deepest images and have $H$ magnitudes and methane colors ranging from 20.8 to 23.0 mag and from 0.23 to $\sim$\,1.0 mag, respectively. Our survey is complete in the magnitude interval $J$\,=\,19--21.7 mag. To complement the near-infrared data, we used the $Z$ (VISTA) and $i'$ (OSIRIS) photometry. We note that neither S\,Ori\,70 nor 73 are detected in the optical images. As observed from S\,Ori\,70 and 73 and field T dwarfs, we expect T-type $\sigma$\,Orionis member candidates to display a very red slope be\-tween the visible and near-infrared wavelengths. Consequently, the following additional photometric criteria were imposed on the newly selected candidates: $i'-J$\,$\ge$\,5 and $Z-J$\,$\ge$\,2.5~mag. It turns out that 5 out of 7 new candidates are clearly detected in the optical images suggesting that they have $i'$\,$\le$\,25 and $Z$\,$\le$\,23~mag, thus discarding them as T-type sources. A sixth candidate is not seen in the optical images, but is detected in the $K'$ (2.12 $\mu$m) frames used in \citet{bihain09}, indicating $J-K'$\,=\,0.96\,mag. This is not a typical color for T dwarfs, leading us to reject it as a methane candidate. Coordinates and photometry of the six rejected candidates are provided in Table~\ref{wrong}. Five sources have brightness beyond the magnitude completeness of our survey. Finder charts are given in Fig.~\ref{charts}. We remark that despite the presence of red methane colors, optical data are critical to the selection of reliable T-type candidates. 

Only one object remains as T dwarf cluster member candidate. It is listed in Table~\ref{newobj} and its finding chart is given in Fig.~\ref{charts}. The candidate lies within the magnitude completeness of our survey.  It appears near a bright source that does not fo\-llow the $\sigma$\,Orionis near-infrared photometric sequence based on its  2MASS data \citep{skrutskie06}. The candidate, S\,Ori\,J053813.2$-$024910, is located at 2\farcs9 southwest of a bright object ($H$\,=\,13.06\,$\pm$\,0.03\,mag) and has a remarkable red $H-CH_{\rm 4off}$ index; its associated photometric uncertainty is do\-mi\-na\-ted by the large error bar of the $CH_{\rm 4off}$-band photometry. We note that our optical data do not have the same spatial re\-so\-lu\-tion as the near-infrared images, and that the wings of the bright sources prevented us from measuring deep $i'$ and $Z$ magnitudes in the immediate su\-rroun\-dings of the stars. Therefore, we cannot conclude that S\,Ori\,J053813.2$-$024910 is a truly faint object complying with our photometric criteria in the red-optical wavelengths. Visual inspection of the $\sigma$\,Orionis {\sl Spitzer} images (3.6, 4.5, 5.8, and 8.0 $\mu$m) obtained by \cite{hernandez07} and \cite{scholz08} yielded that this source is not resolved from the bright star next to it. 

If the new candidate turned out to be a true T-type object, we could estimate its spectral type using $J-H$ and $H-CH_{\rm 4off}$ colors. The object is depicted in Fig.~\ref{cc}. From the position of S\,Ori\,J053813.2$-$024910 in the color-color diagram, we tentatively estimate a classification of $\ge$\,T8\,$\pm$\,1. The candidate has a very red methane color suggesting it might be an object cooler than the T dwarfs. Unfortunately, the $J-H$ color of these kind of ``ultracoool" objects is not known.  \citet{reyle10} provided vo\-lu\-me densities for late-L and T-type dwarfs using astronomical observations of the Canada-France Brown Dwarf Survey. They quoted an object density of 1.4$^{+0.3}_{-0.2}$\,$\times$\,10$^{-3}$\,pc$^{-3}$ for T0--T5.5, 5.3$^{+3.1}_{-2.2}$\,$\times$\,10$^{-3}$\,pc$^{-3}$ for T6 to T8 dwarfs and 8.3$^{+9.0}_{-5.1}$\,$\times$\,10$^{-3}$\,pc$^{-3}$ for dwarfs cooler than T8. From these numbers and the magnitude of our candidate, we derive that the contamination by field $\ge$\,T8 dwarfs in our survey co\-ve\-ring 40.02 arcmin$^2$ (i.e., the deepest HAWK-I images where the candidate was found) is 0.08 objects (19\,$\leq$\,$J$\,$\leq$\,23.2\,mag). The estimated contamination by T dwarfs in the remaining HAWK-I survey is 0.06 (19\,$\leq$\,$J$\,$\leq$\,22.6\,mag). Additional contamination by extragalactic sources may also be present in our survey; \cite{goldman10} discussed that quasars with redshifts between 1.30 and 1.48 have strong H$\alpha$ emission and methane co\-lors that mimic those of T dwarfs. 

Based on the photometrically derived spectral type of S\,Ori\,J053813.2$-$024910, we can discuss whether this candidate can be a $\sigma$\,Orionis member by studying its position in Fig.~\ref{secuencia}. The source de\-via\-tes from the expected sequence. This dis\-cu\-ssion (and the one of previous section) is biased by the known pro\-per\-ties of field T dwarfs, which have a mixture of ages and metallicities; very young T dwarfs with low-gravity atmospheres and same age and metal abundance might display a different behavior. If this candidate were confirmed as a $\sigma$\,Orionis member using follow-up spectroscopic and astrometric data, its mass and surface temperature would be in the range 2--3 M$_{\rm Jup}$ and $\le$1300~K.


\section{Conclusions and final remarks}
We acquired deep HAWK-I (VLT) $J$, $H$, and $CH_{\rm 4off}$ images co\-ve\-ring an area of 119.15 arcmin$^2$ of the young $\sigma$\,Orionis cluster. The completeness magnitudes of these observations are $J$\,=\,21.7 and $H$\,=\,21 mag. The two T-type cluster member candidates previously reported in the literature, S\,Ori\,70 (T5.5\,$\pm$\,1) and 73, are included in the explored region. The ``methane" filter is centered at 1.575 $\mu$m and has a band width of 0.112 $\mu$m. It samples wavelengths bluewards of the $H$-band intense molecular methane absorption. The methane color $H-CH_{\rm 4off}$ has a noticeable dependence on spectral type (or $J-H$), which we calibrated by using the 2MASS photometry of well known stars and brown dwarfs with classifications from F to late-T and by integrating their observed near-infrared spectra available in the literature. The HAWK-I data have allowed us to confirm for the first time the presence of methane absorption in the atmosphere of S\,Ori\,73. Its $J-H$ and $H-CH_{\rm 4off}$ indices are consistent with a spectral type of T4\,$\pm$\,1. Despite being about 0.5--1 mag brighter than S\,Ori\,73, S\,Ori\,70 displays a redder $H-CH_{\rm 4off}$ color compatible with its later spectral classification. 

The new HAWK-I data and VISTA near-infrared images taken as part of the VISTA Orion survey were used to measure the proper motions of both S\,Ori\,70 and 73 by com\-pa\-ri\-son to the previous positions of the objects in near-infrared i\-ma\-ges collected 3.4--7.9 yr ago. We derived $\mu_\alpha\,{\rm cos}\,\delta$\,=\,26.7\,$\pm$\,6.1, $\mu_\delta$\,=\,21.3\,$\pm$\,6.1 mas\,yr$^{-1}$ for S\,Ori\,70, and $\mu_\alpha\,{\rm cos}\,\delta$\,=\,46.7\,$\pm$\,4.9, $\mu_\delta$\,=\,$-$6.3\,$\pm$\,4.7 mas\,yr$^{-1}$ for S\,Ori\,73. These are distinct motions in excess of the Hipparcos $\sigma$\,Orionis proper motion by 4.5-$\sigma$ (S\,Ori\,70) and 7.4-$\sigma$ (S\,Ori\,73). We combined our knowledge of the color and astrometric properties of both T dwarfs to assess their cluster membership in $\sigma$\,Orionis. S\,Ori\,73 has near- and mid-infrared colors undistinguishable from old T3--T5 dwarfs in the field and a large proper motion deviating by 7.4 $\sigma$ from that of the cluster, suggesting that it is probably a field dwarf situated at a distance of 170--200 pc. As reported in the literature, S\,Ori\,70 displays near- and mid-infrared colors redder than typical field T dwarfs, a feature that is qualitatively predicted for low-gravity atmospheres according to state-of-the-art models (see \citet{zapatero08}). We traced back the projected proper motion of S\,Ori\,70 and found that it likely ran close to LDN\,1634 about 6.0$\times$10$^5$ yr ago. LDN\,1634 is a nearby Orion star-forming region located to the southeast of $\sigma$\,Orionis. We conclude that the origin of S\,Ori\,70 remains unclear: it can be a field, foreground mid- to late-T free-floating dwarf with peculiar colors, or an orphan planet ejected through strong dynamical interactions from a nearby Orion star-forming region including $\sigma$ Orionis. If S\,Ori\,70 and 73 were field T4 -- T6 dwarfs we would determine an object volume density a factor of 2\,--\,10 higher than the values of the literature, and the impact in the $\sigma$\,Orionis pla\-ne\-ta\-ry mass function would be dramatic suggesting a decrease in cluster members with planetary masses in the interval 3.5--6 M\,$_{\rm Jup}$.

We also carried out a photometric search for additional T-type cluster member candidates with masses in the interval 2--7 M$_{\rm Jup}$  by combining the HAWK-I data with images collected with OSIRIS ($i'$-band) and the VISTA images ($Z$-band). The search is complete in the magnitude interval $J$\,=\,19--21.7 mag. The fo\-llowing photometric criteria were applied: $J$\,$\ge$\,19, $J-H$\,$\le$\,0.5, $H-CH_{4\rm off}$\,$\ge$\,0.15, $i'-J$\,$\ge$\,5 and $Z-J$\,$\ge$\,2.5~mag, which allowed us to identify candidates with likely spectral types  $\ge$\,T3. We remark that despite the presence of red methane colors, optical data are crutial to selecting reliable T-type candidates because these objects are characterized by extremely red optical-to-near-infrared indices as opposed to (extragalactic) sources that might also show red methane colors. Only one object with $J$\,=\,21.69\,$\pm$\,0.12 mag fulfilled our requirements. However, it lies within a few arcsec from a bright source, a fact that prevented us from studying whether this candidate has genuinely a very red slope from the $i'$ wavelengths to the $J$-band. Its $CH_{\rm 4off}$ color suggests it might have spectral type $\ge$\,T8. Further photometry and spectroscopy are highly required to assess the true nature of this candidate.


\begin{acknowledgements}
We thank E$.$ L$.$ Mart\'\i n and the anonymous referee for constructive discussion on proper motions. We also thank T$.$ Mahoney (IAC) for the English edition of this manuscript. Based on observations made with ESO Telescopes at the Cerro Paranal Observatory under program ID 382.C-0115, and observations made with the Gran Telescopio de Canarias (GTC) operated on the island of La Palma in the Spanish Observatorio del Roque de los Muchachos of the Instituto de Astrof\'\i sica de Canarias. We used the Second Palomar Observatory Sky Survey (POSS-II), which was made by the California Institute of Technology with funds from the National Science Foundation, the National Aeronautics and Space Administration, the National Geographic Society, the Sloan Foundation, the Samuel Oschin Foundation, and the Eastman Kodak Corporation. The Oschin Schmidt Telescope is operated by the California Institute of Technology and Palomar Observatory. This research has been supported by the Spanish Ministry of Science and Innovation (MICINN) under the projects AYA2010-21308-C03-01, AYA2010-21308-C03-02 and AYA2010-20535.
\end{acknowledgements}

\bibliographystyle{aa} 
\bibliography{biblio}

\begin{thebibliography}{62}
\expandafter\ifx\csname natexlab\endcsname\relax\def\natexlab#1{#1}\fi

\bibitem[{{Alves de Oliveira} {et~al.}(2010){Alves de Oliveira}, {Moraux},
  {Bouvier}, {Bouy}, {Marmo}, \& {Albert}}]{alvesdeoliveira10}
{Alves de Oliveira}, C., {Moraux}, E., {Bouvier}, J., {et~al.} 2010, \aap, 515,
  A75+

\bibitem[{{Bally}(2008)}]{bally08}
{Bally}, J. 2008, {Overview of the Orion Complex}, ed. {Reipurth, B.}, 459--+

\bibitem[{{Bally} {et~al.}(2009){Bally}, {Walawender}, {Reipurth}, \&
  {Megeath}}]{bally09}
{Bally}, J., {Walawender}, J., {Reipurth}, B., \& {Megeath}, S.~T. 2009, \aj,
  137, 3843

\bibitem[{{B{\'e}jar} {et~al.}(2001){B{\'e}jar}, {Mart{\'{\i}}n}, {Zapatero
  Osorio}, {Rebolo}, {Barrado y Navascu{\'e}s}, {Bailer-Jones}, {Mundt},
  {Baraffe}, {Chabrier}, \& {Allard}}]{bejar01}
{B{\'e}jar}, V.~J.~S., {Mart{\'{\i}}n}, E.~L., {Zapatero Osorio}, M.~R.,
  {et~al.} 2001, \apj, 556, 830

\bibitem[{{B{\'e}jar} {et~al.}(2011){B{\'e}jar}, {Rebolo}, {Zapatero Osorio},
  {XXXXX}, {XXXXX}, {XXXXX}, {XXXXX}, \& {XXXXX}}]{bejar11}
{B{\'e}jar}, V.~J.~S., {Rebolo}, R., {Zapatero Osorio}, M.~R., {et~al.} 2011,
  submitted

\bibitem[{{B{\'e}jar} {et~al.}(2004){B{\'e}jar}, {Zapatero Osorio}, \&
  {Rebolo}}]{bejar04a}
{B{\'e}jar}, V.~J.~S., {Zapatero Osorio}, M.~R., \& {Rebolo}, R. 2004,
  Astronomische Nachrichten, 325, 705

\bibitem[{{Bihain} {et~al.}(2006){Bihain}, {Rebolo}, {B{\'e}jar}, {Caballero},
  {Bailer-Jones}, {Mundt}, {Acosta-Pulido}, \& {Manchado Torres}}]{bihain06}
{Bihain}, G., {Rebolo}, R., {B{\'e}jar}, V.~J.~S., {et~al.} 2006, \aap, 458,
  805

\bibitem[{{Bihain} {et~al.}(2009){Bihain}, {Rebolo}, {Zapatero Osorio},
  {B{\'e}jar}, {Vill{\'o}-P{\'e}rez}, {D{\'{\i}}az-S{\'a}nchez},
  {P{\'e}rez-Garrido}, {Caballero}, {Bailer-Jones}, {Barrado y Navascu{\'e}s},
  {Eisl{\"o}ffel}, {Forveille}, {Goldman}, {Henning}, {Mart{\'{\i}}n}, \&
  {Mundt}}]{bihain09}
{Bihain}, G., {Rebolo}, R., {Zapatero Osorio}, M.~R., {et~al.} 2009, \aap, 506,
  1169

\bibitem[{{Burgasser} {et~al.}(2002){Burgasser}, {Kirkpatrick}, {Brown},
  {Reid}, {Burrows}, {Liebert}, {Matthews}, {Gizis}, {Dahn}, {Monet}, {Cutri},
  \& {Skrutskie}}]{burgasser02}
{Burgasser}, A.~J., {Kirkpatrick}, J.~D., {Brown}, M.~E., {et~al.} 2002, \apj,
  564, 421

\bibitem[{{Burgess} {et~al.}(2009){Burgess}, {Moraux}, {Bouvier}, {Marmo},
  {Albert}, \& {Bouy}}]{burgess09}
{Burgess}, A.~S.~M., {Moraux}, E., {Bouvier}, J., {et~al.} 2009, \aap, 508, 823

\bibitem[{{Burningham} {et~al.}(2010){Burningham}, {Pinfield}, {Lucas},
  {Leggett}, {Deacon}, {Tamura}, {Tinney}, {Lodieu}, {Zhang}, {Huelamo},
  {Jones}, {Murray}, {Mortlock}, {Patel}, {Barrado Y Navascu{\'e}s}, {Zapatero
  Osorio}, {Ishii}, {Kuzuhara}, \& {Smart}}]{burningham10}
{Burningham}, B., {Pinfield}, D.~J., {Lucas}, P.~W., {et~al.} 2010, \mnras,
  406, 1885

\bibitem[{{Caballero}(2008)}]{caballero08}
{Caballero}, J.~A. 2008, \mnras, 383, 375

\bibitem[{{Caballero} {et~al.}(2007){Caballero}, {B{\'e}jar}, {Rebolo},
  {Eisl{\"o}ffel}, {Zapatero Osorio}, {Mundt}, {Barrado Y Navascu{\'e}s},
  {Bihain}, {Bailer-Jones}, {Forveille}, \& {Mart{\'{\i}}n}}]{caballero07}
{Caballero}, J.~A., {B{\'e}jar}, V.~J.~S., {Rebolo}, R., {et~al.} 2007, \aap,
  470, 903

\bibitem[{{Casali} {et~al.}(2007){Casali}, {Adamson}, {Alves de Oliveira},
  {Almaini}, {Burch}, {Chuter}, {Elliot}, {Folger}, {Foucaud}, {Hambly},
  {Hastie}, {Henry}, {Hirst}, {Irwin}, {Ives}, {Lawrence}, {Laidlaw}, {Lee},
  {Lewis}, {Lunney}, {McLay}, {Montgomery}, {Pickup}, {Read}, {Rees}, {Robson},
  {Sekiguchi}, {Vick}, {Warren}, \& {Woodward}}]{casali07}
{Casali}, M., {Adamson}, A., {Alves de Oliveira}, C., {et~al.} 2007, \aap, 467,
  777

\bibitem[{{Casali} {et~al.}(2006){Casali}, {Pirard}, {Kissler-Patig},
  {Moorwood}, {Bedin}, {Biereichel}, {Delabre}, {Dorn}, {Finger}, {Gojak},
  {Huster}, {Jung}, {Koch}, {Lizon}, {Mehrgan}, {Pozna}, {Silber}, {Sokar}, \&
  {Stegmeier}}]{hawki_2}
{Casali}, M., {Pirard}, J., {Kissler-Patig}, M., {et~al.} 2006, in Presented at
  the Society of Photo-Optical Instrumentation Engineers (SPIE) Conference,
  Vol. 6269, Society of Photo-Optical Instrumentation Engineers (SPIE)
  Conference Series

\bibitem[{{Casewell} {et~al.}(2007){Casewell}, {Dobbie}, {Hodgkin}, {Moraux},
  {Jameson}, {Hambly}, {Irwin}, \& {Lodieu}}]{casewell07}
{Casewell}, S.~L., {Dobbie}, P.~D., {Hodgkin}, S.~T., {et~al.} 2007, \mnras,
  378, 1131

\bibitem[{{Casewell} {et~al.}(2010){Casewell}, {Jameson}, {Burleigh}, {Dobbie},
  {Roy}, {Hodgkin}, \& {Moraux}}]{casewell10}
{Casewell}, S.~L., {Jameson}, R.~F., {Burleigh}, M.~R., {et~al.} 2010, ArXiv
  e-prints

\bibitem[{{Cepa} {et~al.}(2000){Cepa}, {Aguiar}, {Escalera},
  {Gonzalez-Serrano}, {Joven-Alvarez}, {Peraza}, {Rasilla}, {Rodriguez-Ramos},
  {Gonzalez}, {Cobos Duenas}, {Sanchez}, {Tejada}, {Bland-Hawthorn},
  {Militello}, \& {Rosa}}]{cepa00}
{Cepa}, J., {Aguiar}, M., {Escalera}, V.~G., {et~al.} 2000, in Presented at the
  Society of Photo-Optical Instrumentation Engineers (SPIE) Conference, Vol.
  4008, Society of Photo-Optical Instrumentation Engineers (SPIE) Conference
  Series, ed. {M.~Iye \& A.~F.~Moorwood}, 623--631

\bibitem[{{Chiu} {et~al.}(2006){Chiu}, {Fan}, {Leggett}, {Golimowski}, {Zheng},
  {Geballe}, {Schneider}, \& {Brinkmann}}]{chiu06}
{Chiu}, K., {Fan}, X., {Leggett}, S.~K., {et~al.} 2006, \aj, 131, 2722

\bibitem[{{Cushing} {et~al.}(2005){Cushing}, {Rayner}, \& {Vacca}}]{cushing05}
{Cushing}, M.~C., {Rayner}, J.~T., \& {Vacca}, W.~D. 2005, \apj, 623, 1115

\bibitem[{{Dalton} {et~al.}(2006){Dalton}, {Caldwell}, {Ward}, {Whalley},
  {Woodhouse}, {Edeson}, {Clark}, {Beard}, {Gallie}, {Todd}, {Strachan},
  {Bezawada}, {Sutherland}, \& {Emerson}}]{vista_1}
{Dalton}, G.~B., {Caldwell}, M., {Ward}, A.~K., {et~al.} 2006, in Presented at
  the Society of Photo-Optical Instrumentation Engineers (SPIE) Conference,
  Vol. 6269, Society of Photo-Optical Instrumentation Engineers (SPIE)
  Conference Series

\bibitem[{{Emerson} {et~al.}(2006){Emerson}, {McPherson}, \&
  {Sutherland}}]{vista_2}
{Emerson}, J., {McPherson}, A., \& {Sutherland}, W. 2006, The Messenger, 126,
  41

\bibitem[{{Faherty} {et~al.}(2009){Faherty}, {Burgasser}, {Cruz}, {Shara},
  {Walter}, \& {Gelino}}]{faherty09}
{Faherty}, J.~K., {Burgasser}, A.~J., {Cruz}, K.~L., {et~al.} 2009, \aj, 137, 1

\bibitem[{{F{\H u}r{\'e}sz} {et~al.}(2008){F{\H u}r{\'e}sz}, {Hartmann},
  {Megeath}, {Szentgyorgyi}, \& {Hamden}}]{furesz08}
{F{\H u}r{\'e}sz}, G., {Hartmann}, L.~W., {Megeath}, S.~T., {Szentgyorgyi},
  A.~H., \& {Hamden}, E.~T. 2008, \apj, 676, 1109

\bibitem[{{Fritz} {et~al.}(2010){Fritz}, {Gillessen}, {Trippe}, {Ott},
  {Bartko}, {Pfuhl}, {Dodds-Eden}, {Davies}, {Eisenhauer}, \&
  {Genzel}}]{fritz10}
{Fritz}, T., {Gillessen}, S., {Trippe}, S., {et~al.} 2010, \mnras, 401, 1177

\bibitem[{{Geballe} {et~al.}(2002){Geballe}, {Knapp}, {Leggett}, {Fan},
  {Golimowski}, {Anderson}, {Brinkmann}, {Csabai}, {Gunn}, {Hawley},
  {Hennessy}, {Henry}, {Hill}, {Hindsley}, {Ivezi{\'c}}, {Lupton}, {McDaniel},
  {Munn}, {Narayanan}, {Peng}, {Pier}, {Rockosi}, {Schneider}, {Smith},
  {Strauss}, {Tsvetanov}, {Uomoto}, {York}, \& {Zheng}}]{geballe02}
{Geballe}, T.~R., {Knapp}, G.~R., {Leggett}, S.~K., {et~al.} 2002, \apj, 564,
  466

\bibitem[{{Goldman} {et~al.}(2010){Goldman}, {Marsat}, {Henning}, {Clemens}, \&
  {Greiner}}]{goldman10}
{Goldman}, B., {Marsat}, S., {Henning}, T., {Clemens}, C., \& {Greiner}, J.
  2010, \mnras, 405, 1140

\bibitem[{{Golimowski} {et~al.}(2004){Golimowski}, {Leggett}, {Marley}, {Fan},
  {Geballe}, {Knapp}, {Vrba}, {Henden}, {Luginbuhl}, {Guetter}, {Munn},
  {Canzian}, {Zheng}, {Tsvetanov}, {Chiu}, {Glazebrook}, {Hoversten},
  {Schneider}, \& {Brinkmann}}]{golimowski04}
{Golimowski}, D.~A., {Leggett}, S.~K., {Marley}, M.~S., {et~al.} 2004, \aj,
  127, 3516

\bibitem[{{Gonz{\'a}lez Hern{\'a}ndez} {et~al.}(2008){Gonz{\'a}lez
  Hern{\'a}ndez}, {Caballero}, {Rebolo}, {B{\'e}jar}, {Barrado Y
  Navascu{\'e}s}, {Mart{\'{\i}}n}, \& {Zapatero Osorio}}]{hernandez08}
{Gonz{\'a}lez Hern{\'a}ndez}, J.~I., {Caballero}, J.~A., {Rebolo}, R., {et~al.}
  2008, \aap, 490, 1135

\bibitem[{{Hambly} {et~al.}(2008){Hambly}, {Collins}, {Cross}, {Mann}, {Read},
  {Sutorius}, {Bond}, {Bryant}, {Emerson}, {Lawrence}, {Rimoldini}, {Stewart},
  {Williams}, {Adamson}, {Hirst}, {Dye}, \& {Warren}}]{hambly08}
{Hambly}, N.~C., {Collins}, R.~S., {Cross}, N.~J.~G., {et~al.} 2008, \mnras,
  384, 637

\bibitem[{{Herbst} {et~al.}(1999){Herbst}, {Thompson}, {Fockenbrock}, {Rix}, \&
  {Beckwith}}]{herbst99}
{Herbst}, T.~M., {Thompson}, D., {Fockenbrock}, R., {Rix}, H., \& {Beckwith},
  S.~V.~W. 1999, \apjl, 526, L17

\bibitem[{{Hern{\'a}ndez} {et~al.}(2007){Hern{\'a}ndez}, {Hartmann}, {Megeath},
  {Gutermuth}, {Muzerolle}, {Calvet}, {Vivas}, {Brice{\~n}o}, {Allen},
  {Stauffer}, {Young}, \& {Fazio}}]{hernandez07}
{Hern{\'a}ndez}, J., {Hartmann}, L., {Megeath}, T., {et~al.} 2007, \apj, 662,
  1067

\bibitem[{{Hewett} {et~al.}(2006){Hewett}, {Warren}, {Leggett}, \&
  {Hodgkin}}]{hewett06}
{Hewett}, P.~C., {Warren}, S.~J., {Leggett}, S.~K., \& {Hodgkin}, S.~T. 2006,
  \mnras, 367, 454

\bibitem[{{Hodapp} {et~al.}(2009){Hodapp}, {Iserlohe}, {Stecklum}, \&
  {Krabbe}}]{hodapp09}
{Hodapp}, K.~W., {Iserlohe}, C., {Stecklum}, B., \& {Krabbe}, A. 2009, \apjl,
  701, L100

\bibitem[{{Jeffries} {et~al.}(2006){Jeffries}, {Maxted}, {Oliveira}, \&
  {Naylor}}]{jeffries06}
{Jeffries}, R.~D., {Maxted}, P.~F.~L., {Oliveira}, J.~M., \& {Naylor}, T. 2006,
  \mnras, 371, L6

\bibitem[{{Knapp} {et~al.}(2004){Knapp}, {Leggett}, {Fan}, {Marley}, {Geballe},
  {Golimowski}, {Finkbeiner}, {Gunn}, {Hennawi}, {Ivezi{\'c}}, {Lupton},
  {Schlegel}, {Strauss}, {Tsvetanov}, {Chiu}, {Hoversten}, {Glazebrook},
  {Zheng}, {Hendrickson}, {Williams}, {Uomoto}, {Vrba}, {Henden}, {Luginbuhl},
  {Guetter}, {Munn}, {Canzian}, {Schneider}, \& {Brinkmann}}]{knapp04}
{Knapp}, G.~R., {Leggett}, S.~K., {Fan}, X., {et~al.} 2004, \aj, 127, 3553

\bibitem[{{Lawrence} {et~al.}(2007){Lawrence}, {Warren}, {Almaini}, {Edge},
  {Hambly}, {Jameson}, {Lucas}, {Casali}, {Adamson}, {Dye}, {Emerson},
  {Foucaud}, {Hewett}, {Hirst}, {Hodgkin}, {Irwin}, {Lodieu}, {McMahon},
  {Simpson}, {Smail}, {Mortlock}, \& {Folger}}]{lawrence07}
{Lawrence}, A., {Warren}, S.~J., {Almaini}, O., {et~al.} 2007, \mnras, 379,
  1599

\bibitem[{{Lee} {et~al.}(2005){Lee}, {Chen}, {Zhang}, \& {Hu}}]{lee05}
{Lee}, H., {Chen}, W.~P., {Zhang}, Z., \& {Hu}, J. 2005, \apj, 624, 808

\bibitem[{{Lodieu} {et~al.}(2009){Lodieu}, {Zapatero Osorio}, {Rebolo},
  {Mart{\'{\i}}n}, \& {Hambly}}]{lodieu09a}
{Lodieu}, N., {Zapatero Osorio}, M.~R., {Rebolo}, R., {Mart{\'{\i}}n}, E.~L.,
  \& {Hambly}, N.~C. 2009, \aap, 505, 1115

\bibitem[{{Lucas} \& {Roche}(2000)}]{lucas00}
{Lucas}, P.~W. \& {Roche}, P.~F. 2000, \mnras, 314, 858

\bibitem[{{Lucas} {et~al.}(2006){Lucas}, {Weights}, {Roche}, \&
  {Riddick}}]{lucas06}
{Lucas}, P.~W., {Weights}, D.~J., {Roche}, P.~F., \& {Riddick}, F.~C. 2006,
  \mnras, 373, L60

\bibitem[{{Luhman} {et~al.}(2008){Luhman}, {Hern{\'a}ndez}, {Downes},
  {Hartmann}, \& {Brice{\~n}o}}]{luhman08}
{Luhman}, K.~L., {Hern{\'a}ndez}, J., {Downes}, J.~J., {Hartmann}, L., \&
  {Brice{\~n}o}, C. 2008, \apj, 688, 362

\bibitem[{{Marsh} {et~al.}(2010){Marsh}, {Kirkpatrick}, \&
  {Plavchan}}]{marsh10}
{Marsh}, K.~A., {Kirkpatrick}, J.~D., \& {Plavchan}, P. 2010, \apjl, 709, L158

\bibitem[{{McLean} {et~al.}(2003){McLean}, {McGovern}, {Burgasser},
  {Kirkpatrick}, {Prato}, \& {Kim}}]{mclean03}
{McLean}, I.~S., {McGovern}, M.~R., {Burgasser}, A.~J., {et~al.} 2003, \apj,
  596, 561

\bibitem[{{Metchev} {et~al.}(2008){Metchev}, {Kirkpatrick}, {Berriman}, \&
  {Looper}}]{metchev08}
{Metchev}, S.~A., {Kirkpatrick}, J.~D., {Berriman}, G.~B., \& {Looper}, D.
  2008, \apj, 676, 1281

\bibitem[{{Perryman} {et~al.}(1997){Perryman}, {Lindegren}, {Kovalevsky},
  {Hoeg}, {Bastian}, {Bernacca}, {Cr{\'e}z{\'e}}, {Donati}, {Grenon}, {van
  Leeuwen}, {van der Marel}, {Mignard}, {Murray}, {Le Poole}, {Schrijver},
  {Turon}, {Arenou}, {Froeschl{\'e}}, \& {Petersen}}]{perryman97}
{Perryman}, M.~A.~C., {Lindegren}, L., {Kovalevsky}, J., {et~al.} 1997, \aap,
  323, L49

\bibitem[{{Pirard} {et~al.}(2004){Pirard}, {Kissler-Patig}, {Moorwood},
  {Biereichel}, {Delabre}, {Dorn}, {Finger}, {Gojak}, {Huster}, {Jung}, {Koch},
  {Le Louarn}, {Lizon}, {Mehrgan}, {Pozna}, {Silber}, {Sokar}, \&
  {Stegmeier}}]{hawki_1}
{Pirard}, J., {Kissler-Patig}, M., {Moorwood}, A., {et~al.} 2004, in Presented
  at the Society of Photo-Optical Instrumentation Engineers (SPIE) Conference,
  Vol. 5492, Society of Photo-Optical Instrumentation Engineers (SPIE)
  Conference Series, ed. {A.~F.~M.~Moorwood \& M.~Iye}, 1763--1772

\bibitem[{{Rayner} {et~al.}(2009){Rayner}, {Cushing}, \& {Vacca}}]{rayner09}
{Rayner}, J.~T., {Cushing}, M.~C., \& {Vacca}, W.~D. 2009, \apjs, 185, 289

\bibitem[{{Reyl{\'e}} {et~al.}(2010){Reyl{\'e}}, {Delorme}, {Willott},
  {Albert}, {Delfosse}, {Forveille}, {Artigau}, {Malo}, {Hill}, \&
  {Doyon}}]{reyle10}
{Reyl{\'e}}, C., {Delorme}, P., {Willott}, C.~J., {et~al.} 2010, \aap, 522,
  A112+

\bibitem[{{Rosenthal} {et~al.}(1996){Rosenthal}, {Gurwell}, \&
  {Ho}}]{rosenthal96}
{Rosenthal}, E.~D., {Gurwell}, M.~A., \& {Ho}, P.~T.~P. 1996, \nat, 384, 243

\bibitem[{{Sacco} {et~al.}(2008){Sacco}, {Franciosini}, {Randich}, \&
  {Pallavicini}}]{sacco08}
{Sacco}, G.~G., {Franciosini}, E., {Randich}, S., \& {Pallavicini}, R. 2008,
  \aap, 488, 167

\bibitem[{{Saumon} {et~al.}(1996){Saumon}, {Hubbard}, {Burrows}, {Guillot},
  {Lunine}, \& {Chabrier}}]{saumon96}
{Saumon}, D., {Hubbard}, W.~B., {Burrows}, A., {et~al.} 1996, \apj, 460, 993

\bibitem[{{Scholz} \& {Jayawardhana}(2008)}]{scholz08}
{Scholz}, A. \& {Jayawardhana}, R. 2008, \apjl, 672, L49

\bibitem[{{Sherry} {et~al.}(2004){Sherry}, {Walter}, \& {Wolk}}]{sherry04}
{Sherry}, W.~H., {Walter}, F.~M., \& {Wolk}, S.~J. 2004, \aj, 128, 2316

\bibitem[{{Sherry} {et~al.}(2008){Sherry}, {Walter}, {Wolk}, \&
  {Adams}}]{sherry08}
{Sherry}, W.~H., {Walter}, F.~M., {Wolk}, S.~J., \& {Adams}, N.~R. 2008, \aj,
  135, 1616

\bibitem[{{Skrutskie} {et~al.}(2006){Skrutskie}, {Cutri}, {Stiening},
  {Weinberg}, {Schneider}, {Carpenter}, {Beichman}, {Capps}, {Chester},
  {Elias}, {Huchra}, {Liebert}, {Lonsdale}, {Monet}, {Price}, {Seitzer},
  {Jarrett}, {Kirkpatrick}, {Gizis}, {Howard}, {Evans}, {Fowler}, {Fullmer},
  {Hurt}, {Light}, {Kopan}, {Marsh}, {McCallon}, {Tam}, {Van Dyk}, \&
  {Wheelock}}]{skrutskie06}
{Skrutskie}, M.~F., {Cutri}, R.~M., {Stiening}, R., {et~al.} 2006, \aj, 131,
  1163

\bibitem[{{Smith} {et~al.}(2002){Smith}, {Tucker}, {Kent}, {Richmond},
  {Fukugita}, {Ichikawa}, {Ichikawa}, {Jorgensen}, {Uomoto}, {Gunn}, {Hamabe},
  {Watanabe}, {Tolea}, {Henden}, {Annis}, {Pier}, {McKay}, {Brinkmann}, {Chen},
  {Holtzman}, {Shimasaku}, \& {York}}]{smith02}
{Smith}, J.~A., {Tucker}, D.~L., {Kent}, S., {et~al.} 2002, \aj, 123, 2121

\bibitem[{{Spiegel} {et~al.}(2011){Spiegel}, {Burrows}, \&
  {Milsom}}]{spiegel11}
{Spiegel}, D.~S., {Burrows}, A., \& {Milsom}, J.~A. 2011, \apj, 727, 57

\bibitem[{{Tinney} {et~al.}(2005){Tinney}, {Burgasser}, {Kirkpatrick}, \&
  {McElwain}}]{tinney05}
{Tinney}, C.~G., {Burgasser}, A.~J., {Kirkpatrick}, J.~D., \& {McElwain}, M.~W.
  2005, \aj, 130, 2326

\bibitem[{{Zapatero Osorio} {et~al.}(2008){Zapatero Osorio}, {B{\'e}jar},
  {Bihain}, {Mart{\'{\i}}n}, {Rebolo}, {Vill{\'o}-P{\'e}rez},
  {D{\'{\i}}az-S{\'a}nchez}, {P{\'e}rez Garrido}, {Caballero}, {Henning},
  {Mundt}, {Barrado Y Navascu{\'e}s}, \& {Bailer-Jones}}]{zapatero08}
{Zapatero Osorio}, M.~R., {B{\'e}jar}, V.~J.~S., {Bihain}, G., {et~al.} 2008,
  \aap, 477, 895

\bibitem[{{Zapatero Osorio} {et~al.}(2000){Zapatero Osorio}, {B{\'e}jar},
  {Mart{\'{\i}}n}, {Rebolo}, {Barrado y Navascu{\'e}s}, {Bailer-Jones}, \&
  {Mundt}}]{zapatero00}
{Zapatero Osorio}, M.~R., {B{\'e}jar}, V.~J.~S., {Mart{\'{\i}}n}, E.~L.,
  {et~al.} 2000, Science, 290, 103

\bibitem[{{Zapatero Osorio} {et~al.}(2002){Zapatero Osorio}, {B{\'e}jar},
  {Mart{\'{\i}}n}, {Rebolo}, {Barrado y Navascu{\'e}s}, {Mundt},
  {Eisl{\"o}ffel}, \& {Caballero}}]{zapatero02sori70}
{Zapatero Osorio}, M.~R., {B{\'e}jar}, V.~J.~S., {Mart{\'{\i}}n}, E.~L.,
  {et~al.} 2002, \apj, 578, 536

\end{thebibliography}

\end{document}